\documentclass[11pt,a4paper]{article}

\usepackage{bm}
\usepackage{graphicx}
\usepackage{amsmath}
\usepackage{amssymb}
\usepackage[latin1]{inputenc}
\usepackage{chicago} 
\usepackage{latexsym}
\begin{document}

\title{\textbf{Data-driven Algorithms for Dimension Reduction in Causal Inference}}
\author{\textsc{Emma Persson\footnote{Corresponding author. email: emma.persson@stat.umu.se, Phone number: +46907865211 }, Jenny H\"aggstr\"om,} \\ \textsc{Ingeborg Waernbaum and Xavier de Luna} \\
\\
\textit{Department of Statistics, USBE} \\
\textit{Ume{\aa} University} \\
\textit{S-90187 Ume{\aa}, Sweden} \\
}
\date{}
\maketitle

\begin{abstract} 
In observational studies, the causal effect of a treatment may be confounded with variables that are related to both the treatment and the outcome of interest. In order to identify a causal effect, such studies often rely on the unconfoundedness assumption, i.e., that all confounding variables are observed. The choice of covariates to control for, which is primarily based on subject matter knowledge, may result in a large covariate vector in the attempt to ensure that unconfoundedness holds. However, including redundant covariates can affect bias and efficiency of nonparametric causal effect estimators, e.g., due to the curse of dimensionality. 
Data-driven algorithms for the selection of sufficient covariate subsets are investigated. Under the assumption of unconfoundedness the algorithms  search for minimal subsets of the covariate vector. Based, e.g., on the framework of sufficient dimension reduction or kernel smoothing, the algorithms perform a backward elimination procedure assessing the significance of each covariate. Their performance is evaluated in simulations and an application using data from the Swedish Childhood Diabetes Register is also presented.
\end{abstract}



\section{Introduction}

We consider observational studies where the goal is to investigate the causal effect of a treatment on an outcome of interest. In such studies the effect of the treatment may be confounded with other variables that are associated with both the treatment and the outcome of interest. The causal effect of the treatment can be identified if all confounders are observed, which is an assumption commonly referred to as unconfoundedness or no unmeasured confounding. The assumption of unconfoundedness is not testable in general, and thus it must be based on subject matter knowledge. 
In applications where there is a rich set of pretreatment variables, referred to as covariates in the sequel, the unconfoundedness assumption may be more credible. Such applications are nowadays common in the medical and social sciences
due to the increasing possibilities to link administrative and health registers at the individual level. 

This paper proposes and studies methods for data-driven selection of sufficient covariate sets, i.e. sets of covariates such that unconfoundedness holds.  
There may be several different sets of sufficient covariates in any application. Tests for sufficiency of a given subset were proposed by \citeN{JR:97} and further described in a graphical model setting in \citeN{GPR:99}. Although the tests are useful, they require that the empirical researcher defines the specific subset to be tested. 
A directed acyclic graph (DAG) model can also be applied as a basis to evaluate the sufficiency of a covariate set \cite{JP:09}. Using a DAG places great demands on the researcher's knowledge, since all the relations between the observed variables need to be specified. This complete specification is not necessary, and  \citeN{XRW2007} have emphasized conditional independence properties of the variables involved as a guidance for selection of sets (see also \citeN{VdW2011}). A large part of the literature is concerned with the balancing property \cite{RR1983,DR:97}. 
This corresponds to identifying covariates whose distribution differs between treated and untreated, e.g., selecting relevant covariates for a propensity score model. 
\citeN{brookhart2006}, \citeN{BK:11} and  \citeN{vansteelandt2012} study covariate selection in a parametric setting, where the association of the covariates with both the treatment and the outcome is considered. 
In such a parametric setting a variance reduction of the estimator is obtained even when the covariates included are associated with outcome and not with the treatment assignment. 
 For a semi-parametric estimator, an inverse probability weighted (IPW) estimator, \citeN{LD:04} describe the variance reduction  when adding a covariate in the propensity score model that is solely related to outcome.
In similar contexts, different methods for simultaneous covariate selection and model fitting have been proposed by \citeN{vdLG:10}, \citeN{hill2012bayesian}  and \citeN{MRM:04}.

 In general, the covariate set will have an influence on both the large and small sample properties of an estimator. 
 For estimators of the average causal effect under unconfoundedness, using a subset containing all covariates predicting the outcome has advantages when it comes to efficiency \cite{XRW2007,WL:11}. 
 However, knowledge of the reduced subset discarding covariates related to the outcome and not to the treatment, is sometimes necessary to reach a lower efficiency bound \cite{H:04}. 
 
For nonparametric  estimators, bias typically dominates variance, where the former dramatically depends on the dimension of the covariate set, e.g. \cite{AI2006}. Therefore, it is important to keep the cardinality of the covariate vector as low as possible.
 \citeN{XRW2007}  developed a theory for selection of minimal subsets of covariates which are sufficient for unconfoundedness, and proposed two general algorithms for covariate selection.
In this paper, we build on this theory by proposing data-driven algorithms for the selection of sufficient covariate sets. In the case of continuous covariates we study the use of marginal co-ordinate hypothesis tests \cite{cook2004,Lcooknach2005} based on the theory of sufficient dimension reduction in regression \cite{cook1994,cook1996}, to find sufficient subsets. When discrete covariates are present, as it is often the case in applications, we study the use of a kernel smoothing method \cite{HLR:04,HLR:07b,LRW:09}. Other model-free covariate selection methods could be used to implement the algorithms. The approach is here to select covariates without making strong model assumptions with the final aim to estimate a causal effect nonparametrically, e.g. using matching \cite{RR1983,AI2006}. However, for comparison, we also implement the algorithms using parametric models for the outcome and treatment in combination with AIC and LASSO \cite{FHT:10} selection.

We study the finite sample properties of the algorithms, where continuous, discrete and mixed continuous-discrete sets of covariates are considered. 
 In particular, the properties of matching estimators and an IPW estimator, are studied when sufficient covariate subsets are selected with the described algorithms. For instance, smaller mean squared errors (MSE) are achieved when using our algorithms, than when using all the covariates predicting the treatment assignment. In general, decreasing cardinality of the covariate set when possible yields better results.
Covariate selection in the context of record linkage studies is illustrated in an application where we estimate the effect of low compulsory school grades on acute complications of Type 1 Diabetes Mellitus.

The paper is organized as follows. 
In the next section the theoretical framework and the covariate selection procedure are introduced. In Section 3, results from a simulation study are presented and an application concerning the effect of low compulsory school grades on acute complications of Type 1 Diabetes Mellitus can be found in Section 4. A discussion concludes the paper.

\section{Covariate selection: context, theory and algorithms}
\subsection{Context}

We consider a binary treatment, $T$, which will take on the value of 1 if treatment is received and 0 otherwise. 
For each unit we define two potential outcomes \cite{Neyman23,Rubin74,Rubin77}, $Y_1$ if the unit is treated and $Y_0$ if the unit is untreated. Only one of the potential outcomes can be observed for each unit, and we denote the observed response $Y$, where $Y=TY_1+(1-T)Y_0$. Let $X$ denote a set of covariates observed for all units. The parameter of interest considered is the average treatment effect,
\begin{equation}
\beta = \text{E}(Y_1 - Y_0),
\end{equation}
although the results presented below are useful for other summaries of the distribution of $Y_1 - Y_0$.
In observational studies, where treatment assignment is not randomized, unconfoundedness, when it holds, allows us to identify causal parameters such as $\beta$. 
Consider the following assumptions,
\begin{description}
\item[A.1] [unconfoundedness] $\hspace{0.5cm} Y_t \perp\hskip -7pt \perp T\mid  {X}$,  $\hspace{0.5cm}  t = 0, 1$,
\item[A.2] [positivity] $ \hspace{0.5cm}  \text{P}(T=t\mid  {X})>0$, $\hspace{0.5cm}  t = 0, 1$.
\end{description}
Under A.1 and A.2, the average treatment effect can be identified since 
$
\beta = \text{E}(Y_1 - Y_0) = \text{E}[ \text{E}(Y_1 \mid T=1,  {X}) - \text{E}(Y_0 \mid T=0,  {X}) ].
$
In this paper, we refer to assumption A.1 when discussing unconfoundedness, sometimes referred to as weak unconfoundedness \cite{Imbens2000}. In situations where A.1 and A.2 hold the parameter $\beta$ may be estimated nonparametrically by conditioning on the covariates $X$, e.g., using matching and/or IPW estimators; see, e.g., \citeN{Imbens2009} for a review on estimation of causal effects under unconfoundedness.

\subsection{Theory}
Under the assumption that unconfoundedness holds for a full covariate set $X$, subsets of $X$ can be defined such that the treatment and the potential outcomes are independent given these subsets. 
Consider starting by selecting a subset consisting of covariates affecting the treatment, or alternatively a subset affecting the outcome. Among the covariates in these two, possibly different, subsets there may be variables that are not necessary, i.e., covariates that are not necessary for A.1 to hold. 
Therefore, we can consider reducing these subsets further.
Moreover, these reduced subsets may be different depending on the initial subsets from which the covariates were selected.

The subsets mentioned above are now formally defined through their conditional independence properties. The defined sets below exist and are unique under mild assumptions (see Appendix \ref{cov} and \citeN{XRW2007}):

\begin{description}
	\item[] 
\textbf{Definition 1} [Covariates predicting treatment] Let $ {X}_T$ be defined as the set $ {X}_T \subseteq  {X}$ of minimum cardinality such that $\text{P}(T \mid  {X}) = \text{P}(T \mid  {X}_T)$.
	\item[] 
\textbf{Definition 2} [Confounders among covariates predicting treatment] Let $ {Q}_t$ be defined as the set $ {Q}_t \subseteq  {X}_T$ of minimum cardinality such that $\text{P}(Y_t \mid  {X}_T) = \text{P}(Y_t \mid  {Q}_t)$, for $t$ = 0, 1.
	\item[] 
\textbf{Definition 3} [Covariates predicting outcome] Let $ {X}_t$ be defined as the set $ {X}_t \subseteq  {X}$ of minimum cardinality such that $\text{P}(Y_t \mid  {X}) = \text{P}(Y_t \mid  {X}_t)$, $t$ = 0, 1.
	\item[] 
\textbf{Definition 4} [Confounders among covariates predicting outcome] Let $ {Z}_t$ be defined as the set $ {Z}_t \subseteq  {X}_t$ of minimum cardinality such that $\text{P}(T \mid  {X}_t) = \text{P}(T \mid  {Z}_t)$, $t$ = 0, 1.
\end{description}

\begin{figure}\label{fig}
\begin{center}
\ifx\JPicScale\undefined\def\JPicScale{1}\fi
\unitlength \JPicScale mm
\begin{picture}(155.59,35)(0,40)
\put(143.89,40.5){\makebox(0,0)[cc]{$Y_0$}}

\put(130.51,62.73){\makebox(0,0)[cc]{$V_3$}}

\linethickness{0.3mm}
\multiput(130.51,59.09)(0.12,-0.18){85}{\line(0,-1){0.18}}
\put(140.71,43.74){\vector(2,-3){0.12}}
\linethickness{0.3mm}
\multiput(147.06,43.74)(0.12,0.22){71}{\line(0,1){0.22}}
\put(147.06,43.74){\vector(-1,-2){0.12}}
\put(155.59,62.73){\makebox(0,0)[cc]{$V_4$}}

\put(42.73,63.39){\makebox(0,0)[cc]{}}

\put(77.92,90){\makebox(0,0)[cc]{}}

\put(3.93,66.5){\makebox(0,0)[cc]{Panel a)}}

\put(88.72,66.36){\makebox(0,0)[cc]{Panel b)}}

\put(10.52,39.63){\makebox(0,0)[cc]{$T$}}

\put(38.08,39.63){\makebox(0,0)[cc]{$Y_0$}}

\put(24.3,62.26){\makebox(0,0)[cc]{$V_2$}}

\put(0,63){\makebox(0,0)[cc]{$V_1$}}

\put(48.42,62.26){\makebox(0,0)[cc]{$V_3$}}

\linethickness{0.3mm}
\multiput(0.19,57.73)(0.12,-0.24){57}{\line(0,-1){0.24}}
\put(7.08,44.16){\vector(1,-2){0.12}}
\linethickness{0.3mm}
\multiput(13.97,44.16)(0.12,0.24){57}{\line(0,1){0.24}}
\put(13.97,44.16){\vector(-1,-2){0.12}}
\linethickness{0.3mm}
\multiput(27.75,57.73)(0.12,-0.24){57}{\line(0,-1){0.24}}
\put(34.64,44.16){\vector(1,-2){0.12}}
\linethickness{0.3mm}
\multiput(41.53,44.16)(0.12,0.24){57}{\line(0,1){0.24}}
\put(41.53,44.16){\vector(-1,-2){0.12}}
\put(140.71,43.74){\makebox(0,0)[cc]{}}

\put(0,45.5){\makebox(0,0)[cc]{}}

\put(148.07,36.87){\makebox(0,0)[cc]{}}

\put(134.69,59.09){\makebox(0,0)[cc]{}}

\linethickness{0.3mm}
\multiput(88.72,59.09)(0.12,-0.18){85}{\line(0,-1){0.18}}
\put(98.91,43.74){\vector(2,-3){0.12}}
\linethickness{0.3mm}
\multiput(105.27,43.74)(0.12,0.22){71}{\line(0,1){0.22}}
\put(105.27,43.74){\vector(-1,-2){0.12}}
\linethickness{0.3mm}
\put(116.8,62.73){\line(1,0){9.53}}
\put(126.33,62.73){\vector(1,0){0.12}}
\put(144.89,40.1){\makebox(0,0)[cc]{}}

\put(101.25,40.91){\makebox(0,0)[cc]{$T$}}

\put(88.72,62.73){\makebox(0,0)[cc]{$V_1$}}

\put(113.79,62.73){\makebox(0,0)[cc]{$V_2$}}

\end{picture}
\end{center}
\caption{Two examples illustrated by  DAGs}
\label{ex}
\end{figure}
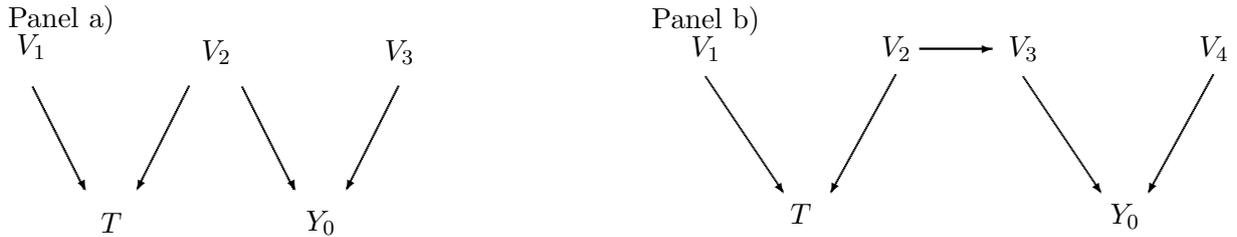

To illustrate the definitions, consider the examples displayed in Figure \ref{ex} for two DAGs. In panel a), where $X$ consists of three covariates, we have that $X_T = \{ V_1, V_2\}$, $Q_0 = \{ V_2\}$, $X_0 = \{ V_2, V_3\}$ and $Z_0 = \{ V_2\}$. In panel b), we have four covariates that reduce to $X_T = \{ V_1, V_2\}$, $Q_0 = \{ V_2\}$, $X_0 = \{ V_3, V_4\}$ and $Z_0 = \{ V_3\}$.
In the second panel, we see that the reduced sets may differ depending on if the relation between the covariates and the treatment, or the outcome, is considered first.

Under assumption A.1 we have that, for $t$ = 0, 1, $Y_t  \perp\hskip -7pt \perp T \mid  {X}_T$, $Y_t  \perp\hskip -7pt \perp T \mid  {Q}_t$, $Y_t  \perp\hskip -7pt \perp T \mid  {X}_t$ and $Y_t \perp\hskip -7pt \perp T \mid  {Z}_t$ \cite[Prop. 3 \& 4]{XRW2007}, i.e. unconfoundedness holds when conditioning on the reduced sets. 
Under mild assumptions (see Appendix \ref{cov}) all the sets defined are identified \cite[Prop. 9 \& 10]{XRW2007}, and $ {Q}_1$, $ {Q}_0$, $ {Z}_1$ and $ {Z}_0$ are minimal in the sense that they cannot be reduced without violating unconfoundedness \cite[Prop. 7]{XRW2007}.

\subsection{Algorithms}

Based on the results of identification and minimality of the defined subsets \cite{XW2005,XRW2007} two covariate identification algorithms are available, see Table \ref{T1}.
A reduction of the initial covariate set is achieved, in a two step identification process of removing redundant covariates, by evaluating the conditional independence implied by the definitions of the subsets  $X_T$, $Q_t$, $X_t$ and $Z_t$, $t$ = 0, 1.
Algorithm A first removes covariates that are conditionally independent of the treatment given the rest of the covariates. In a second step covariates conditionally independent of the potential outcomes in each treatment group given the remainder of the covariates are excluded.  
In algorithm B, the order of the process is reversed. First, we remove covariates conditionally independent of the potential outcomes in each treatment group given the rest. Secondly,  covariates conditionally independent of the treatment given the remaining covariates are excluded.

\begin{table}
\caption{Identification of reduced covariate sets.}
\centering
\begin{tabular}{ll}
\hline
&\textbf{Algorithm A}: Identification of subsets $ {Q}_0$ and $ {Q}_1$ \\
\hline
Step 1: & Identify $ {X}_T$ such that $T  \perp\hskip -7pt \perp  {X} \setminus  {X}_T \mid  {X}_T$ holds.  \\
Step 2: & For $t$ = 0, 1  \\     
        & Identify $ {Q}_t \subseteq  {X}_T$ such that $Y_t  \perp\hskip -7pt \perp  {X}_T \setminus  {Q}_t \mid  {Q}_t, T = t$ holds.  \\ \\
\hline
&\textbf{Algorithm B}: Identification of subsets $ {Z}_0$ and $ {Z}_1$ \\
\hline
& For $t$ = 0, 1 \\ 
Step 1:         & Identify $ {X}_t$ such that $Y_t  \perp\hskip -7pt \perp  {X} \setminus  {X}_t \mid  {X}_t, T = t$ holds. \\
Step 2: & Identify $ {Z}_t \subseteq  {X}_t$ such that $T  \perp\hskip -7pt \perp  {X}_t \setminus  {Z}_t \mid  {Z}_t$ holds.  \\ 
\hline
\end{tabular}
\label{T1}
\end{table}

 Next, we propose a data-driven implementation of the identification algorithms based on model-free evaluation of the conditional independence statements.
In the following subsections we describe two approaches, subject to different conditions. 
For a covariate set containing solely continuous covariates we propose a method based on sufficient dimension reduction (SDR) \cite{cook1994,cook1996} and marginal co-ordinate hypothesis (MCH) tests \cite{Lcooknach2005} (Section  \ref{SDR}). 
When the covariate set includes discrete valued variables we propose an approach based on kernel smoothing, see Section \ref{NP}.

\subsubsection{Testing of conditional independence with SDR.} \label{SDR}

SDR is originally a graphical tool in regression analysis. For the regression problem, with a $p$-dimensional vector $X$ and a response $U$, the goal is to characterize the conditional distribution of $U|X$. SDR aims to reduce the dimension of $X$ by replacing it with a minimal set of linear combinations of $X$ with no loss in information about $U|X$ and with no parametric model initially assumed. 
More formally, let $P_{\mathcal{S}}$ be the orthogonal projection operator onto the subspace $\mathcal{S}$, then the linear combinations form a subspace $\mathcal{S} \subseteq \mathbb{R}^p$ such that 
\begin{equation}
U  \perp\hskip -7pt \perp  {X} \mid P_{\mathcal{S}}  X.
\label{sub} 
\end{equation}
All subspaces that satisfy \eqref{sub} are called dimension reduction subspaces and when, in turn, the intersection of all of these subspaces satisfy \eqref{sub} it is called the central subspace, $\mathcal{S}_{U|X}$. The central subspace exists and is unique and minimal under some conditions \cite{cook1994,cook1996}.

In order to apply dimension reduction in the algorithms of Table \ref{T1}, SDR in terms of variable selection is now considered. The goal is to identify components of $X$ that do not affect treatment and/or the potential outcomes.
Let $X$, a vector of continuous variables, be partitioned as $ {X} = ( {X}_1^T,  {X}_2^T)^T$ where $ {X}_1$ includes $p_1$ variables of $X$, with $0 \leq p_1 \leq p$, and $ {X}_2$ includes the remaining $p_2 = p - p_1$ elements, such that 
\begin{equation}
U  \perp\hskip -7pt \perp  {X}_2 \mid  {X}_1.
\label{rel}
\end{equation}
Let also the columns of the $p \times k$ matrix $ {\eta}$ be a basis for the central subspace $\mathcal{S}_{U|X}$ where $k$ is the dimension of $\mathcal{S}_{U|X}$, and partition $ {\eta} = ( {\eta}_1^T,  {\eta}_2^T)$ according to the partition of $X$. Then if \eqref{rel} holds, $ {X}_2$ will have no information about $U$, given $ {X}_1$, and so $ {\eta}_2 = 0$. These rows of zero vectors in the basis correspond to the set of targeted variables, $ {X}_2$. This can be more formally stated as 
\begin{equation}
P_\mathcal{H} \mathcal{S}_{U|X} = \mathcal{O}_p,
\label{H0}
\end{equation}
where $\mathcal{H} = span[(0,  {I}_{p_2})^T]$ is the subspace corresponding to the co-ordinates $ {X}_2$ and 
$\mathcal{O}_p$ indicates the origin in $\mathbb{R}^p$. 
The relationship in \eqref{rel} corresponds directly to each step in Algorithm A and B, where $U$ corresponds to treatment or the potential outcome.

 \citeN{Lcooknach2005} implement \eqref{rel} in the form of a model-free backward elimination procedure to reduce the dimension of the variable vector $X$. 
The procedure start with all variables present and then eliminates the least significant variable until no more variables can be removed. The test used to examine the significance of a variable is 
a MCH test which evaluates the conditional independence in \eqref{rel} where $ {X}_2$ is one specifically targeted covariate and $ {X}_1$ is the rest of the covariates, i.e., the null and alternative hypotheses are 
\begin{equation}
\textnormal{H}_0: P_\mathcal{H} \mathcal{S}_{U|X} = \mathcal{O}_p \quad \mbox{versus} \quad \textnormal{H}_1: P_\mathcal{H} \mathcal{S}_{U|X} \neq \mathcal{O}_p.
\label{hyp}
\end{equation}
In practice, the procedure requires construction of $\mathcal{H}$ and also estimation of $\mathcal{S}_{U|X}$. Let us say that the first variable in $X$ is the target variable. In that case $\mathcal{H} = span[(1, 0, ...,0)^T]$. The construction of $\mathcal{H}$ is similar for other target variables in $X$. 
When estimating the central subspace, methods such as sliced inverse regression (SIR) \cite{li1991} and sliced average variance estimation \cite{SCW:07} can be implemented. 
For details about the test statistic, for the MCH test, and its asymptotic distribution see \citeN{cook2004}.

\subsubsection{Assessing conditional independence with kernel smoothing.}\label{NP}
The second approach of assessing the conditional independency is based on kernel smoothing \cite{HLR:04,HLR:07b,LRW:09}. 
In contrast to MCH testing this method allows a mix of continuous and discrete covariates. The method relies on generalized kernels and have the property that irrelevant covariates are "smoothed out" if the bandwidths are selected by cross-validation \cite{HLR:04,HLR:07b}. Consider the model for the response
$$
U_i=g( {X}_i)+\epsilon_i,\,\,i=1,\ldots,n
$$
\noindent
with $g(\cdot)$ an unknown smooth function, $ {X}_i$ the covariate vector of individual $i$ and $\epsilon_i$ a zero-mean, finite variance error term. 
Here, interactions may also be considered by including them as separate covariates.
Furthermore, suppose that the covariate vector can be partitioned as $ {X}_i=( {X}_{ic}^T,  {X}_{id}^T)^T$, where $ {X}_{ic}$ is a $q\times 1$ vector containing the continuous covariates and $ {X}_{id}$ is  a $r\times 1$ vector containing the categorical covariates. Let $w(\cdot)$ denote a univariate kernel function for the continuous covariates, then the product kernel function is defined as
$$
W_h(x_i^c,x_j^c)=\prod_{k=1}^q\frac{1}{h_k}w\bigg(\frac{x_{ik}^c-x_{jk}^c}{h_k}\bigg),
$$
where $x_{ik}^c$ and $x_{jk}^c$ are the $i$:th and $j$:th observations of the $k$:th continuous covariate and $h_k \in [0,\infty)$ is the corresponding bandwidth. For the categorical covariates, supposing that the covariates in $ {X}_{id}$ are arranged such that the $r_o$ ordered (i.e., ordinal) covariates are followed by the $r_u$ unordered (i.e., nominal) covariates ($r_o+r_u=r$), we define the product kernel function as
$$
L_{\lambda}(x_i^d,x_j^d)=\bigg[\prod_{l=1}^{r_o}\big(\lambda_l\big)^{|x_{il}^d-x_{jl}^d|}\bigg]\bigg[\prod_{m=r_o+1}^r\big(\lambda_m\big)^{I(x_{im}^d\neq x_{jm}^d)}\bigg],
$$
where $x_{il}^d$ and $x_{jl}^d$ are the $i$:th and $j$:th observations of the $l$:th categorical covariate and $\lambda_l \in [0,1]$ is the corresponding bandwidth. Similarly,  $x_{im}^d$ and $x_{jm}^d$ are the $i$:th and $j$:th observations of the $m$:th categorical covariate and the corresponding bandwidth is $\lambda_m\in [0, (c_m^u-1)/c_m^u]$, with $c_m^u$ the number of categories of the $m$:th categorical covariate. $I(A)$ denotes the indicator function that assumes the value 1 if $A$ is true and 0 otherwise.
Combining the above product kernel functions results in the following kernel estimator of $g( {X}_i)$,
\begin{equation}
\hat{g}( {X}_i;  {h},\lambda)=\frac{\sum_{j=1}^nW_h(x_i^c,x_j^c)L_{\lambda}(x_i^d,x_j^d)U_j}{\sum_{j=1}^nW_h(x_i^c,x_j^c)L_{\lambda}(x_i^d,x_j^d)}.
\label{npcme}
\end{equation}
\noindent
The bandwidths, $( {h}^T, \lambda^T)^T=(h_1,\ldots,h_q, \lambda_1, \ldots, \lambda_r)^T$,  are selected by cross-validation, i.e., by minimizing
$$
CV( {h},\lambda)=\frac{1}{n}\sum_{i=1}^n\big(U_i-\hat{g}^{-i}( {X}_i;  {h},\lambda)\big)^2,
$$
where $\hat{g}^{-i}( {X}_i;  {h},\lambda)$ is the estimate of $g( {X}_i)$ based on data where the $i$:th observation is left out. Covariates are "smoothed out" when their bandwidths are large, i.e., close to the maximum bandwidth values.

We estimate the subsets in the algorithms in Table \ref{T1} by, in each step, regressing the prescribed outcome variable on the prescribed covariates using the estimator in \eqref{npcme}. Exclusion of a covariate is determined by the bandwidth value of the covariate, for a threshold value chosen by the analyst. If the bandwidth value is larger than the threshold value it is taken as an indication that the outcome variable is conditionally independent of the covariate.

The covariate selection is not sensitive to the choice of threshold for continuous covariates, an arbitrarily large value, e.g., 100, can be chosen since the bandwidths for irrelevant and relevant covariates are well separated \cite{HLR:04,HLR:07b}. The choice of threshold for discrete covariates may, however, have a larger impact on the final covariate selection.
To investigate the sensitivity of the final covariate selection, with respect to the chosen thresholds, the following action is recommended: For each covariate in each set in the algorithms we have access to the bandwidths selected by cross-validation. These can be inspected to see if there are any covariates that were close to being included or excluded in a certain set. Taking this information into consideration, the analyst can then re-run the selection with altered thresholds, if needed. 
%


\section{Simulation}
A simulation study is performed to evaluate the implementation of the covariate selection algorithms and 
to investigate the finite sample properties of three non/semi-parametric estimators using the selected subsets.

\subsection{Simulation design}
Ten covariates, $ {X} = (X_1, ..., X_{10})$, are generated with three covariate distribution setups; all continuous, all discrete and a mixture of continuous and discrete covariates.
In the continuous covariate case, the covariates are standard normally distributed, and in the discrete covariate case all covariates are generated from a Bernoulli distribution with success probability 0.5. 
The mixed covariate scenario, include three covariates, $X_2, X_4$ and $X_5$, generated from a standard normal distribution and the remaining seven from a Bernoulli distribution with success probability 0.5.
The treatment variable, $T$, is generated from $n$ Bernoulli trials with the treatment probability
\begin{equation}
\text{P}(T=1 \mid  {X}) = \frac{1}{1 + \exp(a_0 -kX_1 -kX_2 -kX_3 -kX_4 -kX_7)},
\label{T}
\end{equation}
where the coefficient $k$ is equal to 1 if the corresponding covariate is continuous and 2 if it is discrete. 
The intercept, $a_0$, in model \eqref{T} is chosen so that E($T$) = 0.5, i.e., $a_0=0$ in the continuous covariate case, $a_0=5$ in the discrete covariate case and $a_0=3$ in the mixed covariate case.

Four types of outcome models are generated, one linear, one binary and two different nonlinear models yielding a constant and non constant treatment effect, respectively. 
The linear potential outcome models are
\begin{equation}
\begin{split} 
Y_0 &= 2+ 2^kX_1+ 2^kX_2 + 2^kX_5+ 2^kX_6 + 2^kX_8 + \varepsilon_0 = 2+ f(X_1,X_2,X_5,X_6,X_8) + \varepsilon_0, \\ 
Y_1 &= 4+ 2^kX_1+ 2^kX_2 + 2^kX_5+ 2^kX_6 + 2^kX_8  + \varepsilon_1= 4+ f(X_1,X_2,X_5,X_6,X_8) + \varepsilon_1,
\end{split}
\label{c1}
\end{equation}
where $\varepsilon_0$ and $\varepsilon_1$ are independent and standard normally distributed.
The binary outcome model is generated from $n$ Bernoulli trials with outcome probabilities
\begin{equation}
\begin{split} 
\text{P}(Y_0 =1 \mid  {X})&= \{1 + \exp[b_0 + f(X_1,X_2,X_5,X_6,X_8) ]\} ^{-1}, \\ 
\text{P}(Y_1 =1 \mid  {X})&= \{1 + \exp[b_1 + f(X_1,X_2,X_5,X_6,X_8) ]\} ^{-1},
\end{split}
\label{c2}
\end{equation}
where $f(\cdot)$ is defined in \eqref{c1}. In the continuous covariates scenario $(b_0, b_1) = (2,4)$ and $ - f(\cdot)$ is used, and for the discrete and mixed covariate setup $(b_0, b_1) = (5,7)$ and 
$(b_0, b_1) = (-2,-4)$, respectively.
The more complex nonlinear potential outcome models, for the continuous covariates scenario, are 
\begin{equation}
\begin{split} 
Y_0 &= 2 + \frac{7X_6}{0.5 + (X_1 + 2)^2} + (2+2j)X_2  + 2X_5 + 2X_8 + \varepsilon_0, \\ 
Y_1 &= 4 + \frac{(7+ 2j)X_6}{0.5 + (X_1 + 2)^2} + 2X_2  + 2X_5 + 2X_8 + \varepsilon_1,
\end{split}
\label{c345}
\end{equation}
where $j=0, 1$. 
If discrete covariates are present the nonlinear potential outcome models are specified as, 
\begin{equation}
\begin{split} 
Y_0 &= 2 - \frac{6X_6}{\log[(X_1 + 1.4)^2]}                 + |  [2^k + 3j(k-1)]  X_2  + 2^k X_5 | + 2^kX_8 + \varepsilon_0, \\ 
Y_1 &= 4 - \frac{(6 + 3j)X_6}{\log[(X_1 + 1.4)^{2 + j}]}  + 2^k | X_2  + X_5 | + 2^kX_8 + \varepsilon_1.
\end{split}
\label{d345}
\end{equation}
The treatment effect is constant (non constant) for $j=0$ ($j=1$).
The coefficient, $k$, in the models \eqref{T}, \eqref{c1}, \eqref{c2} and \eqref{d345}, is equal to 1 if the corresponding covariate is continuous and 2 if it is discrete. 

In all simulation scenarios the covariates are independent except for $X_7$ and $X_8$ that have a correlation of 0.7.
For model \eqref{c345} and \eqref{d345}, a second covariate structure is also used where additional correlation between covariates is added. Here, in addition to the correlation between  $X_7$ and $X_8$, we also have Corr($X_1$, $X_2$) = Corr($X_5$, $X_6$) = 0.5 and Corr($X_3$, $X_9$) = Corr($X_4$, $X_9$) = 0.25.
For the all simulation scenarios with 10 covariates, the target covariate subsets are
$
 {X}_T = \{X_1,X_2,X_3,X_4,X_7\}$,
$ {Q}_0 =  {Q}_1 = \{X_1,X_2,X_7\}$, 
$ {X}_0 =  {X}_1 = \{X_1,X_2,X_5,X_6,X_8\}$, and
$ {Z}_0 =  {Z}_1 = \{X_1,X_2,X_8\} 
$.

To study a scenario with more covariates, we also investigate a design  with additional covariates added. To the first mixed covariate setup, we add 10 more covariates, eight discrete covariates from a Bernoulli distribution with success probability varying between 0.35 and 0.70. The additional discrete covariates are independent of the potential outcomes and treatment assignment but are correlated with each other: Corr($X_{11}$, $X_{12}$) = Corr($X_{15}$, $X_{16}$) = 0.5 and Corr($X_{13}$, $X_{19}$) = Corr($X_{14}$, $X_{19}$) = 0.25.
Two normally distributed covariates ($X_{17}, X_{18}$) with mean and variance equal to one, are also added. These continuous covariates have a correlation of 0.8 and are related to the potential outcomes and the treatment assignment in the following way:
\begin{equation*}
\text{P}(T=1 \mid  {X}) = \frac{1}{1 + \exp(4.5 -2X_1 -2X_2 -X_3 -X_4 -2X_7 -2X_{17}) },
\label{T20cov}
\end{equation*}
and
\begin{equation*}
\begin{split} 
Y_0 &= 1 + 4X_1  + 3 | X_2 + 0.42 |  +  X_5 +  X_6 + 4X_8 + 3 | X_{18}  - 0.66 | + \varepsilon_0, \\ 
Y_1 &= 5 + 4X_1   - 3 | X_2 - 0.42  |  +  X_5 +  X_6 + 4X_8  - 3 | X_{18}  - 1.34 | + \varepsilon_1.
\end{split}
\label{Y20cov}
\end{equation*}
For this simulation scenario, with 20 covariates, the target covariate subsets include one additional covariate,
$
 {X}_T = \{X_1,X_2,X_3,X_4,X_7,X_{17}\}$,
$ {Q}_0 =  {Q}_1 = \{X_1,X_2,X_7,X_{17}\}$, 
$ {X}_0 =  {X}_1 = \{X_1,X_2,X_5,X_6,X_8,X_{18}\}$, and
$ {Z}_0 =  {Z}_1 = \{X_1,X_2,X_8,X_{18}\} 
$.

\subsection{Covariate Selection}

The target subsets are, in the sequel, denoted by $\mathsf{S}$.
In order to fulfill the assumption of unconfoundedness, $(Y_1,Y_0) \perp\!\!\!\perp T \mid {\mathsf{S}}$, a selected subset ($\hat{\mathsf{S}}$) has to include  $X_1$, $X_2$ and at least one from the set of the correlated confounders $\{X_7,X_{8}\}$. For the  simulation scenario with 20 covariates at least one from the set $\{X_{17},X_{18}\}$ also needs to be included to uphold the condition. 

When we have only continuous covariates the algorithms are implemented with MCH tests with 10\% significance level in a backward stepwise procedure. We use SIR to estimate the central subspace.
The underlying assumptions for MCH testing, primarily associated with the distribution of $X$, 
are not violated in the designs, see  \citeN{Lcooknach2005} and the references therein.

If discrete covariates are present, we use kernel smoothing for assessing the conditional independency and the bandwidth thresholds are 0.5 of the maximum bandwidth for the binary covariates and 100 for the continuous covariates. The selection of the continuous covariates is not sensitive to the chosen threshold, since the bandwidths for relevant and irrelevant covariates are well separated and not near the chosen value. The binary covariates are more sensitive to the choice of threshold, since the bandwidths lie in the range $[0, 0.5]$. Here, the threshold 0.25 is chosen and minor variations in this value would not affect the selection.
Although, there is interaction between covariates in  \eqref{d345}, no interaction terms are considered in the model for the selection procedure.
In addition, the assumption of additive error terms is violated for the binary outcome model \eqref{c2}.


All simulations are repeated with 1000 iterations each, with sample sizes $n$ = 500, 1000. Data generation and all  computations are performed with the software \texttt{R} \cite{R} and subset selection is performed with the \texttt{R}-package \texttt{CovSel} \cite{JSS}.

For comparison, we select covariates in each step of the algorithms using two commonly used methods: model selection with AIC and LASSO \cite{FHT:10}. 
We use a  backward stepwise approach with AIC, where we assume a linear model without interaction terms. The function \texttt{stepAIC}, available in the  \texttt{MASS} package \cite{MASS}, is used for the computations. 
For the LASSO method we fit a generalized linear model using a logit link function for the binary outcomes.
We use a large number of predictors for the purpose of approximating an unknown regression function. The model considers all two-way interactions and power exponents 1, 2 and 3, as well as -1 and -2 for corresponding continuous covariates. In the case of 10 continuous covariates this yields 95 predictors in total. 
In the second step of the algorithms, the model again considers all predictors mentioned above for all covariates that are, in some way, included among the chosen predictors in the first step. 
Cross-validation, using the function \texttt{cv.glmnet}, in the  \texttt{glmnet} package  \cite{FHT:10}, is used to select the LASSO regularization parameter (\texttt{lambda.1se}).
The selected subsets, $\hat{\mathsf{S}}$, only include the related covariates, i.e., no interactions or higher order terms are further considered in the propensity score model.

\subsection{Estimation of the average treatment effect}

We investigate the performance of an IPW estimator \cite{LD:04} and two matching estimators \cite{AI2006}, with separate  matching criteria, based on the selected covariate subsets. 

The matching estimator performs one-to-one matching with replacement, i.e., matched pairs where the matched unit may be used as a match more than once. We start by defining the matching criteria between two units $i$ and $i'$ from opposite treatment groups.
 The first matching criterion is the  Euclidian vector norm, 
\begin{equation}\label{vecnorm}
\lVert \hat{ {S}}_i - \hat{ {S}}_{i'} \rVert = \sqrt{(\hat{ {S}}_i - \hat{ {S}}_{i'})^T (\hat{ {S}}_i - \hat{ {S}}_{i'})},
\end{equation}
where $\hat{ {S}}_i$ and $\hat{ {S}}_{i'}$ are the selected covariate vectors for unit $i$ and its potential match $i'$. 
We may also match on the propensity score, thus another matching criterion is 
\begin{equation}\label{propscore}
\vert \hat{e}(\hat{S}_i) - \hat{e}(\hat{S}_{i'})   \vert ,
\end{equation}
where $\hat{e}(\hat{S}_i)$ is the estimated propensity score value for unit $i$. 
The propensity score is estimated by 
\begin{equation*}
\hat{e}(\hat{S}) = \hat{\text{P}}(T=1 \mid \hat{ {S}}) = \frac{1}{1 + \exp[ - (\hat{\delta}_0 + \hat{ {\delta}}^T \hat{ {S}} )]},
\end{equation*}
where $ \hat{\delta}$, a maximum likelihood estimator, is a column-vector of the same dimension as $\hat{ {S}}$.
For \eqref{vecnorm} and \eqref{propscore} separately, let $Y_{i'}(i)$ be the observed outcome of unit $i'$ that minimizes the matching criterion for unit $i$, i.e., the matched unit's observed outcome.
The matching estimator of the average treatment effect is 
\begin{equation}
\hat{\beta}_{m} = \frac{1}{n} \sum_{i=1}^n (2T_i - 1)[ Y_{i} - Y_{i'}(i) ],
\label{sm}
\end{equation}
both when matching on the vector norm or when matching on the propensity score.

In addition to the matching estimators, we apply a simple IPW estimator \cite{LD:04},
\begin{equation}
\hat{\beta}_{ipw} = \frac{1}{n} \sum_{i=1}^n \left[ \frac{T_iY_i}{\hat{e}(\hat{S})} - \frac{(1-T_i) Y_i}{1-\hat{e}(\hat{S})} \right],
\label{ipw}
\end{equation}
controlling for the selected sets.

\subsection{Simulation results}


\begin{figure}
\begin{center}
\includegraphics[scale=0.6]{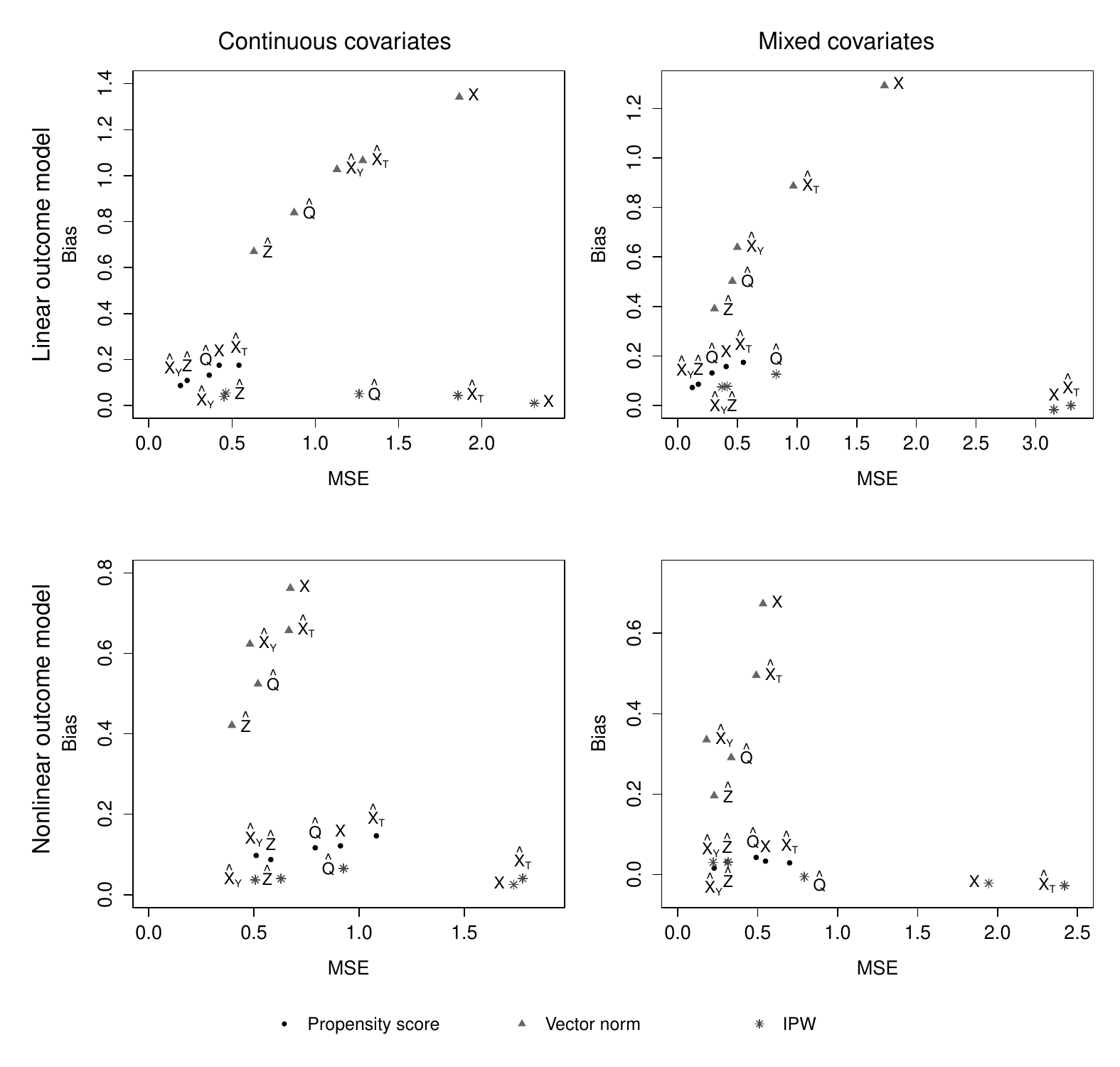}
\end{center}
\caption{Bias and MSE results from two simulation setups with 10 covariates with sample size 500. Conditioning on the full covariate set $X$, covariates predicting the treatment ${\hat{X}_{\text{\tiny{T}}}}$, the union of the minimal subsets from algorithm A, $\hat{ Q}$, the union of the covariates predicting the outcome, ${\hat{X}_{\text{\tiny{Y}}}}$, and the union of minimal subsets from algorithm B, $\hat{Z}$. }
\label{fig}
\end{figure}

The results from the covariate selection algorithms are summarized in Table \ref{selectCont}, \ref{selectDisc} and \ref{selectMix} in Appendix \ref{tables}, where we give selection success rates. Three definitions of success are used for the selected subset, $\hat{\mathsf{S}}$; i) unconfoundedness holds, $(Y_1,Y_0) \perp\!\!\!\perp T \mid \hat{\mathsf{S}}$, ii) the target subset is included in the selected subset ($\mathsf{S}  \subseteq \hat{\mathsf{S}}$), and iii) equal subsets ($\mathsf{S}  = \hat{\mathsf{S}}$).
The results show that, in all simulation setups and for both implementation methods, the selected subsets are generally sufficient for upholding the assumption of unconfoundedness, with a proportion higher than 92.4\% for a sample size of 1000. 
The success rates for when the selected subsets include their target subsets are lower due to the correlated covariates, $X_7$ and $X_8$, being interchanged. Redundant covariates are at times selected lowering the success rate for equal subsets.

In comparison, the AIC method breaks down for the nonlinear outcome model scenarios (see Table \ref{selectMix20}), but has high success rates in the discrete covariate distribution setup where the nonlinearity is not as severe.
The LASSO performs equivalently with the kernel smoothing method but does not, to the same extent, find $X_1$ in the nonlinear outcome model scenarios compared to the MCH method. 
For the full result tables from the covariate selection algorithms using AIC  and LASSO contact the authors.

The bias and MSE from the estimation of $\beta$, using the estimators \eqref{sm} and \eqref{ipw}, are summarized  in Figure \ref{fig} for two of the simulation setups.
The estimators are evaluated when controlling for the full covariate set, $X$, and the selected subsets $\hat{X}_T$, $\hat{Q} = \hat{ {Q}}_0 \cup \hat{ {Q}}_1$, $\hat{ {X}}_Y = \hat{ {X}}_0 \cup \hat{ {X}}_1$ and  $\hat{ {Z}} = \hat{ {Z}}_0 \cup \hat{ {Z}}_1$.
For all simulation scenarios, we see that when matching on the vector norm \eqref{vecnorm}, conditioning on the full covariate set, $X$, yields the largest bias. This is largely due to the cardinalities of the selected subsets which are considerably smaller. This large bias results in large MSE, even though the variance is small, for the estimator controlling for $X$.
For the IPW estimator, \eqref{ipw}, or when matching is performed on the propensity score, \eqref{propscore}, the differences in bias between the estimators utilizing the full covariate set and the selected covariate sets are not as pronounced. 
We can see that the largest variance and MSE can be found when conditioning on the full covariate set $X$ or $\hat{ {X}}_T$; the set of covariates commonly used in propensity score methods.

For all three approaches, estimators based on the selected set $\hat{ {Q}}$ yield lower MSE than $\hat{ {X}}_T$. In turn, the MSE when utilizing $\hat{ {Q}}$ is commonly larger than that resulting from using subsets from algorithm B. Exceptions may be found for the vector norm matching estimator where bias sometimes dominates variance. 
Since $\hat{ {X}}_Y$ usually yields the smallest variance of all selected subsets, but not necessarily smallest bias, the smallest MSE can, in all simulation setups, be found for one of the two subsets $\hat{ {Z}}$ and $\hat{ {X}}_Y$.
The results show that, for the chosen simulation setups, algorithm B is preferable over algorithm A. 

Table \ref{estimateCont}, \ref{estimateDisc} and  \ref{estimateMix} in Appendix \ref{tables}, display bias, standard deviation and MSE for the matching estimator \eqref{sm} and the IPW estimator  \eqref{ipw}, as well as the median cardinality of the conditioning sets, for the sample size 500. 
The results from the estimators based on all the different sets of covariates show a decrease in bias and variance as sample size increases to 1000, but exhibit a similar pattern as for the smaller sample size, and are therefore omitted.

Finally, we show the results from the simulation setup with additional covariates, 20 covariates in total. In Table \ref{selectMix20} we see that when the initial covariate set is larger the methods perform similarly.
However, the kernel smoothing method is computationally heavy and the running time increased notably. 
The selection using LASSO have the highest success rates, for the smaller sample size, when it comes to selecting sets that uphold unconfoundedness, although  performs equivalently to the  kernel smoothing method for $n=1000$. The  AIC performs poorly due to the nonlinearity in the outcome model and  often exclude vital confounders.
In Table \ref{estimateMix20} we see results from the estimation of the average treatment effect, $\beta$.
 For $n=500$, the conditioning sets ${\hat{ {Q}}}$ and ${\hat{ {Z}}}$ do not satisfy unconfoundedness  6.8\% and 9.8\% of the times, respectively, using the kernel smoothing method.
For the AIC method the respective proportions are much larger with 76.4\% and 59.8\% failure.
The poor selection affects the estimation of $\beta$, more specifically resulting in a larger bias when using the selected subsets from the AIC selection. 
This is more prominent for the propensity score matching and  the IPW estimator. 
When matching on the vector norm the cardinality of the conditioning set, per se, also increases the bias.
When matching on the propensity score, using the selected covariate sets from the kernel smoothing approach gives the smallest MSE, compared to LASSO and AIC.
For the IPW estimator we see a larger variance, which dominates the bias, and hence a smaller MSE when using the AIC method. 
Using LASSO results in a large conditioning set, since many redundant covariates are often selected.
 The increased cardinality affects the performance of the estimators and we can often see the largest MSE for the LASSO despite the high success rates selecting sets that uphold unconfoundedness.

\begin{table}
\caption{The proportion (\%) of times the selected subset, $\hat{\mathsf{S}}$, satisfies three conditions of unconfoundedness. Simulation scenario with 20 covariates with mixed (discrete and continuous) covariate distributions and a nonlinear outcome model. }
\centering
\begin{tabular}{cllccccccc}
 \hline
 & Selection &   Satisfied & \multicolumn{3}{c}{$\mathsf{S}$ in Algorithm A} &  \multicolumn{4}{c}{ $\mathsf{S}$ in Algorithm B} \\ 
$n$ & method &  condition & $ {X}_{\text{\tiny{T}}}$ & $ {Q}_0$ & $ {Q}_1$& $ {X}_0$ & $ {X}_1$ & $ {Z}_0$ & $ {Z}_1$ \\
 \hline
500 & Kernel   & $Y_t \perp\!\!\!\perp T \mid \hat{\mathsf{S}}$	& 94.0 & 87.8 & 88.0 & 100.0 & 100.0 & 82.2 & 83.7 \\ 
  &  smoothing  & $\mathsf{S}  \subseteq \hat{\mathsf{S}}$ & 42.9 & 50.9 & 50.6 & 41.6 & 42.2 & 54.9 & 57.0 \\ 
   &  & $\mathsf{S}=\hat{\mathsf{S}}$ & 2.1 & 9.3 & 7.8 & 9.4 & 8.2 & 13.3 & 13.5 \\ 
 \cline{2-10}
& LASSO   & $Y_t \perp\!\!\!\perp T \mid \hat{\mathsf{S}}$	& 100.0 & 98.3 & 99.1 & 100.0 & 99.8 & 99.8 & 99.8 \\ 
   &   & $\mathsf{S}  \subseteq \hat{\mathsf{S}}$ & 99.5 & 62.7 & 58.4 & 99.9 & 99.3 & 60.2 & 61.3 \\ 
     & & $\mathsf{S}=\hat{\mathsf{S}}$ & 0.0 & 0.2 & 0.6 & 0.1 & 1.0 & 0.2 & 1.5 \\ 
 \cline{2-10}
    &  AIC  & $Y_t \perp\!\!\!\perp T \mid \hat{\mathsf{S}}$	& 100.0 & 8.1 & 9.2 & 17.1 & 15.8 & 17.1 & 15.8 \\ 
    & & $\mathsf{S}  \subseteq \hat{\mathsf{S}}$ & 99.3 & 6.4 & 6.6 & 10.1 & 9.3 & 6.2 & 7.2 \\ 
    & & $\mathsf{S}=\hat{\mathsf{S}}$ & 7.2 & 2.6 & 1.7 & 0.8 & 0.6 & 1.7 & 1.6 \\ 
 \hline
1000 & Kernel   & $Y_t \perp\!\!\!\perp T \mid \hat{\mathsf{S}}$	& 100.0 & 98.3 & 99.3 & 100.0 & 100.0 & 96.7 & 96.8 \\ 
   &   & $\mathsf{S}  \subseteq \hat{\mathsf{S}}$ & 55.6 & 67.4 & 68.6 & 64.8 & 65.0 & 68.4 & 69.5 \\ 
    &  & $\mathsf{S}=\hat{\mathsf{S}}$ & 7.6 & 14.6 & 15.0 & 17.9 & 19.2 & 16.4 & 15.9 \\ 
 \cline{2-10}
 & LASSO   & $Y_t \perp\!\!\!\perp T \mid \hat{\mathsf{S}}$	& 100.0 & 99.9 & 100.0 & 100.0 & 100.0 & 100.0 & 100.0 \\ 
    &   & $\mathsf{S}  \subseteq \hat{\mathsf{S}}$ & 100.0 & 54.4 & 66.8 & 100.0 & 100.0 & 62.0 & 59.8 \\ 
      & & $\mathsf{S}=\hat{\mathsf{S}}$ & 0.0 & 0.0 & 0.9 & 0.0 & 0.9 & 0.6 & 2.1 \\ 
 \cline{2-10}
    &   AIC  & $Y_t \perp\!\!\!\perp T \mid \hat{\mathsf{S}}$	& 100.0 & 10.8 & 10.3 & 17.9 & 19.9 & 17.9 & 19.9 \\ 
      & & $\mathsf{S}  \subseteq \hat{\mathsf{S}}$ & 100.0 & 8.5 & 8.2 & 10.3 & 11.3 & 6.3 & 5.9 \\ 
     & &  $\mathsf{S}=\hat{\mathsf{S}}$ & 9.4 & 2.4 & 2.5 & 0.5 & 0.4 & 1.9 & 1.7 \\ 
   \hline
\end{tabular}
\label{selectMix20}
\end{table}

\begin{table}
\caption{Results from the mixed covariate distribution setup, with 20 covariates,  when estimating $\beta$ using matching on the vector norm and the propensity score and IPW. $m$ is the median cardinality of the conditioning set.}
\centering
\begin{tabular}{cllcccccccccc}
 \hline
& Selection &  & \multicolumn{3}{c}{Vector norm} & \multicolumn{3}{c}{Propensity Score}  & \multicolumn{3}{c}{IPW} &  \\ 
$n$ & method   &  Set       & Bias  & SD & MSE   & Bias  & SD & MSE & Bias  & SD & MSE & $m$  \\
 \hline
 500 & 	 & $ {X}$ 					& 2.778 & 0.347 & 7.838		& 0.577 & 0.936 & 1.209 		& 0.374 & 4.077 & 16.763	&  20\\
	 \cline{2-13}
 & Kernel   & ${\hat{ {X}}_{\text{\tiny{T}}}}$& 2.346 & 0.452 & 5.707 	& 0.619 & 0.940 & 1.267		& 0.518 & 3.350 & 11.490		 &  9\\
 &  smoothing&  ${\hat{ {Q}}}$  		& 1.848 & 0.548 & 3.717 		& 0.504 & 0.820 & 0.926		& 0.618 & 2.450 & 6.386		 &  6\\  
 & & ${\hat{ {X}}_{\text{\tiny{Y}}}}$ 		& 2.095 & 0.417 & 4.562 		& 0.375 & 0.679 & 0.601		& 0.539 & 2.216 & 5.200		 &  8\\ 
 &  &${\hat{ {Z}}}$	 				& 1.780 & 0.493 & 3.412 		& 0.444 & 0.713 &  0.705		& 0.592 & 2.211 & 5.239		 &  5\\ 
 \cline{2-13}
 &  LASSO& ${\hat{ {X}}_{\text{\tiny{T}}}}$	& 2.552 & 0.386 & 6.663		& 0.500 & 0.934 & 1.122 		& 0.438 & 3.573 & 12.957		 &  14\\
 &  &  ${\hat{ {Q}}}$  					& 2.403 & 0.463 & 5.986 		& 0.456 & 0.888 & 0.997		& 0.460 & 3.313 & 11.189		 &  13\\
 & & ${\hat{ {X}}_{\text{\tiny{Y}}}}$ 			& 2.717 & 0.358 & 7.512 		& 0.542 & 0.887 & 1.081		& 0.435 & 3.803 & 14.656		 &  19\\
 &  &${\hat{ {Z}}}$	 					& 2.446 & 0.440 & 6.178 		& 0.489 & 0.904 & 1.056		& 0.466 & 3.595 & 13.145		 &  14\\
 \cline{2-13}
 &  AIC & ${\hat{ {X}}_{\text{\tiny{T}}}}$	& 2.333 & 0.418 & 5.618 		& 0.570 & 0.938 & 1.205 		& 0.415 & 4.097 & 16.960		 &  8\\ 
 &  &  ${\hat{ {Q}}}$  				& 1.387 & 0.493 & 2.166 		& 0.893 & 0.635 & 1.201		& 0.951 & 1.206 &  2.360			 &  5\\ 
 & & ${\hat{ {X}}_{\text{\tiny{Y}}}}$ 		& 2.045 & 0.452 & 4.387 		& 0.738 & 0.606 & 0.913		& 0.825 & 1.376 &  2.575		 &  10\\ 
 &  &${\hat{ {Z}}}$	 				& 1.438 & 0.478 & 2.296		& 0.728 & 0.617 & 0.910		& 0.856 & 1.264 &  2.331		 &  5\\ 
 \hline
 1000 & 	 & $ {X}$ 					& 2.617 & 0.227 & 6.900		& 0.426 & 0.731 & 0.716 		& 0.305 & 3.492 & 12.285	&  20\\
	 \cline{2-13}
 & Kernel   & ${\hat{ {X}}_{\text{\tiny{T}}}}$& 1.902 & 0.371 & 3.757 	& 0.376 & 0.732 & 0.678		& 0.316 & 4.241 & 18.088		 &  7\\
 &  smoothing&  ${\hat{ {Q}}}$  		& 1.435 & 0.430 & 2.245 		& 0.324 & 0.642 & 0.517		& 0.502 & 2.437 & 6.191		 &  5\\  
 & & ${\hat{ {X}}_{\text{\tiny{Y}}}}$ 		& 1.760 & 0.346 & 3.219 		& 0.280 & 0.513 & 0.342		& 0.487 & 1.584 & 2.747		 &  7\\ 
 &  &${\hat{ {Z}}}$	 				& 1.436 & 0.413 & 2.233 		& 0.291 & 0.525 &  0.360		& 0.505 & 1.554 & 2.670		 &  5\\ 
 \cline{2-13}
 &  LASSO& ${\hat{ {X}}_{\text{\tiny{T}}}}$	& 2.344 & 0.305 & 5.589		& 0.377 & 0.730 & 0.675 		& 0.221 & 4.526 & 20.534		 &  14\\
 &  &  ${\hat{ {Q}}}$  					& 2.204 & 0.390 & 5.012 		& 0.344 & 0.699 & 0.607		& 0.348 & 4.280 & 18.440		 &  13\\
 & & ${\hat{ {X}}_{\text{\tiny{Y}}}}$ 			& 2.551 & 0.261 & 6.577 		& 0.303 & 0.707 & 0.662		& 0.378 & 3.463 & 12.133		 &  19\\
 &  &${\hat{ {Z}}}$	 					& 2.227 & 0.381 & 5.103 		& 0.380 & 0.697 & 0.630		& 0.331 & 4.247 & 12.143		 &  13\\
 \cline{2-13}
  & AIC & ${\hat{ {X}}_{\text{\tiny{T}}}}$	& 2.040 & 0.335 & 4.275 		& 0.422 & 0.742 & 0.730 		& 0.333 & 3.503 & 12.380		 &  8\\ 
 &  &  ${\hat{ {Q}}}$  				& 1.258 & 0.421 & 1.759 		& 0.764 & 0.579 & 0.919		& 0.898 & 1.583 &  3.314			 &  5\\ 
 & & ${\hat{ {X}}_{\text{\tiny{Y}}}}$ 		& 1.886 & 0.390 & 3.710 		& 0.645 & 0.552 & 0.720		& 0.800 & 1.241 &  2.179		 &  10\\ 
 &  &${\hat{ {Z}}}$	 				& 1.272 & 0.406 & 1.782		& 0.632 & 0.536 & 0.687		& 0.803 & 1.220 &  2.133		 &  5\\ 
 \hline
\end{tabular}
\label{estimateMix20}
\end{table}

\section{Effect of school achievements on acute complications of type 1 diabetes mellitus }

Type 1 Diabetes Mellitus (T1DM) is a complex autoimmune disease typically diagnosed in childhood or adolescence and can lead to deadly complications. For children with this disease it is important to regulate the  blood sugar levels. As the individuals become older this task largely falls on the individuals themselves. Poor management of T1DM can cause acute complications such as ketoacidosis or coma. Considering the school achievements as a proxy for the individuals ability to manage their disease we investigate the effect of low compulsory school grades on acute complications of T1DM. We compare the estimated effects when controlling for the original set of covariates and the subsets of covariates obtained by the algorithms described in Section 2.
In the Swedish Childhood Diabetes Register (SCDR), all children 0-15 years of age diagnosed with T1DM in Sweden are registered since 1977 \cite{diab}. By linkage to the Longitudinal Integration Database for Health Insurance and Labour Market Studies, the Inpatient Register, the Swedish Register of Education and the Multigenerational Register, a range of socioeconomic, demographic, and other variables are available for the children as well as their parents.

The study population comprises of all children diagnosed with T1DM before the age of 15 and that received a compulsory school grade in Sweden between 1991 and 1997. Due to missing information on the mother and/or the father, 10 individuals are excluded from the study.
The outcome is a binary variable defined as equal to 1 if the individual has been hospitalized with a main or secondary diagnosis of T1DM with ketoacidosis, hyperosmolarity, lactic acidosis, hypoglycaemic coma and/or coma. The outcome, i.e., hospitalization, is measured after the grade is received and in the 10 years following. The 24 individuals that died before the end of this 10 year period are not included in the study, resulting in a sample size of 2234.
The compulsory school grades in Sweden, at this time, are measured on a scale from one to five, five being the highest mark and the mean grade is computed as an overall measure.
Individuals are defined as treated if their received mean grade, after the ninth year of compulsory school, is lower than the 20th percentile, in this case a mean grade lower than 2.6.
The possible observed confounders are; gender, age at onset of T1DM and socioeconomic characteristics of the parents. The socioeconomic variables are measured one year prior to receiving the grade, see Table \ref{desc} in Appendix \ref{tables} for a description.

The algorithms in Table \ref{T1} are applied to the set of 22 covariates and the selection is performed using the kernel smoothing based method described in Section \ref{NP}. 
Bandwidth thresholds are 100 for the continuous covariates (age, income) and 0.5 for the discrete ordered covariates (educational level). For the binary covariates a smaller threshold of 0.25 of the maximum bandwidth is chosen, since many of the covariates are believed to be a measure of the same characteristic, i.e., socioeconomic status. The cardinality of the chosen sets only increases with one or two for some of the selected sets if the threshold is increased to 0.5.
Computations are performed with the software \texttt{R} \cite{R} and the package \texttt{CovSel} \cite{JSS} is used for subset selection. 
The software allows for specification of variables that should be retained in the selected subsets regardless of the results of the assessment of their significance. However, from discussions with diabetes scientists, none of the observed covariates are known confounders prior to the analysis. Thus, no such restriction on the chosen sets are made.
The following subsets are selected,
\begin{eqnarray*}
&\hat{ {X}}_T &= \{ageM, soc, educM, educF,  sickM, sickF, ampolM, ampolF, erM, retM \}, \\
&\hat{ {Q}}_0 &=  \{ageM, educM\}, \\
&\hat{ {Q}}_1 &= \{soc \}, \\
&\hat{ {X}}_0 &=    \{ageM, single, educM,  sickM, ampolF, erM, injM \}, \\
&\hat{ {X}}_1 &= \{ageM, soc, single,  unempM, unempF, retF, injM, injF \}, \\
&\hat{ {Z}}_0 &=  \{ageM, educM,  ampolF, erM, injM   \}, \\
&\hat{ {Z}}_1 &=  \{ageM, soc, unempF, injM, injF  \}.
\end{eqnarray*}
The union of the corresponding subscripted subsets are thereupon used to control for confounding in the estimation of $\beta$. Covariates that are not selected for any of the subsets include; gender, age at onset of T1DM, the fathers age and early retirements benefits, as well as both parents salary income. 
The subset $\hat{ {X}}_T$ includes 10 covariates, where the parents unemployment status and occupational injury annuity, and the fathers retirement pension and single mother status, have not been included.  
The subset $\hat{ {Q}} = \hat{ {Q}}_0 \cup \hat{ {Q}}_1$ only includes three covariates; the mothers age, social benefits and educational level. From algorithm B, the subset $\hat{ {X}}_Y = \hat{ {X}}_0 \cup \hat{ {X}}_1$ includes 12 covariates, the excluded covariates are $ampolM$, $retM$, $educF$ and $sickF$. An additional, $single$, $sickM$, $unempM$ and $retF$ have been removed in the subset $\hat{ {Z}} = \hat{ {Z}}_0 \cup \hat{ {Z}}_1$, which contains eight covariates in total. 
All selected subsets achieve a substantial reduction of the initial covariate vector. The amount and type of covariates chosen for the sets differs, which may be due to the correlation between them.

\begin{table}
\caption{Estimated average treatment effects ($\hat{\beta}$), standard deviations (SD) and confidence intervals (CI).}
\centering
\begin{tabular}{lcccc}
\hline
Conditioning covariate set & Number of covariates  &  $\hat{\beta}$ & SD & 95\% CI \\
\hline
All Variables 					&	22			& 0.079	&  0.031	& (0.017, 0.140) \\ 
Algorithm A, $\hat{ {X}}_T$ 		&	10		 	& 0.077	&  0.029	&  (0.019, 0.135)  \\
Algorithm A, $\hat{ {Q}}$ 			&	3			& 0.058	&  0.023	&  (0.012, 0.103)   \\
Algorithm B, $\hat{ {X}}_Y$ 		&	12			& 0.062	&  0.030	&   (0.003, 0.120)   \\
Algorithm B, $\hat{ {Z}}$ 			&	8			& 0.055	&  0.025	&   (0.006, 0.105)   \\
\hline
\end{tabular}
\label{effect}
\end{table}

The matching estimator defined in \eqref{sm}, is used to estimate the effect. Matching is performed on the propensity score, estimated by \eqref{propscore} and the standard deviation is calculated taking the matching procedure into account \cite{AI2006}.
The average treatment effect, $\hat{\beta}$, can be found in Table \ref{effect}.
We see that, compared to the unadjusted difference in mean which is 0.082, the effect of having low compulsory school grades on acute complications of T1DM is smaller when controlling for confounding covariates. 
The positive effect is significant regardless of the set of covariates we choose to control for. However, it should be noted that the covariate selection procedure is not accounted for in the assessment of the uncertainty.
Finally, we see that the estimated effect and variance is smaller when controlling for the minimal subsets $\hat{ {Q}}$ and $\hat{ {Z}}$.

\section{Discussion}
In this paper, we present data-driven algorithms for the selection of covariate subsets sufficient for estimating causal effects. We implement the algorithms using both MCH testing for continuous covariates and kernel smoothing for discrete and mixed covariate sets.
We consider a binary treatment, although the covariate selection procedure can also be applied to treatments with more than two levels, $t \in \{1, ..., c\}$, if the steps in the algorithm are adjusted accordingly.

 In a simulation study, the proposed covariate selection algorithms are evaluated, and compared with LASSO and AIC selection.
 The finite sample properties of an IPW and two matching estimators are investigated utilizing the selected subsets.  
 The results show that the algorithms perform well in selecting sufficient covariate subsets and in improving the resulting mean squared error of the estimators. It should be noted here that the underlying conditions for the validity of the MCH testing hold for the data generating mechanisms chosen.

The dimension of the initial covariate set may have an impact on the performance of the algorithms. 
The kernel smoothing approach is not suitable for a large initial covariate set, where most of the covariates are relevant for the unconfoundedness assumption to hold. However, if a large majority of the covariates are in fact irrelevant,  \citeN{HLR:04,HLR:07b}  show that  they  will asymptotically be smoothed out. 
In addition, although the latter theoretical results consider uncorrelated covariates, simulations studies have shown that the performance of the kernel smoothing method is not sensitive to correlated covariates \cite{HLR:07b,LOR:09}. Note that with many covariates and observed units, the kernel smoothing approach is computationally heavy.
Finally, situations where the number of covariates is larger than the sample size are not considered in this paper. However, there are methods based on sufficient dimension reduction for such settings that can be used for covariate selection \cite{LY:08}.


For both the matching and IPW estimators we typically see a decrease in MSE when reducing the dimension from the full set, where redundant covariates are included, to one of the selected subsets.
Exceptions can be found for the propensity score based estimators, where MSE is often largest when using all covariates predicting the treatment ($X_T$).
Based on previous results \cite{WL:11,H:04,LD:04} and the simulations of this paper, we do not recommend using $X_T$, even though it is commonly the covariate set that would be selected for a propensity score model.

When the initial set consists of covariates predicting the treatment, excluding covariates that are not predicting the outcome will  result in a smaller MSE. Moreover, when the initial set consists of covariates related to the outcome, the change in MSE when reducing the dimension will depend on the relation between the bias reduction and the variance increase. 
For simulated designs, matching or weighting on the propensity score has the lowest MSE when using all covariates predicting the outcome. However, the propensity scores used are correctly specified. If the propensity score specification is unknown, higher cardinality of the covariate set is expected to be more costly in terms of MSE. This can be seen when matching on the vector norm, for which our results show that MSE is decreased also when excluding covariates solely related to the outcome, so that sets with lowest cardinality perform best.

The use of the covariate selection algorithms is also illustrated in a case study where we estimate the effect of low compulsory school grades on acute complications of T1DM by reducing the cardinality of the covariate set matched for. We find a significant effect that individuals with low grades are more likely to be hospitalized with acute complications of T1DM. Viewing the grades as a marker for the individual's ability to self-regulate the disease, this effect may be of interest to policy makers, since, educational measures can be directed to this group.

It is generally acknowledged that inference ignoring the uncertainty due to the covariate selection may be overoptimistic. Model averaging is a solution, which is natural within a Bayesian framework (e.g.,  \citeN{ZD:14}). Model averaging has also been proposed and studied within a frequentist mode of inference in \citeN{CH:2003}, see also  \citeN{vansteelandt2012}, using asymptotic approximations. This approach is based on the availability of a class of parametric models including the true data generating mechanisms. Generalizing such results to our model-free covariate selection setting is a challenging and important direction for future research.

Finally, covariate selection is part of the design of an observational study and as such should be outcome-free \cite{LR2008} in the sense that one should not be able to gear selection towards a given desired result (causal effect). This is guaranteed when using the algorithms presented herein since covariate selection is performed by either focusing on $Y_0\mid T=0$ or $Y_1\mid T=1$, that is, a causal effect is never estimated. This is not the case for outcome model-based methods.

\section*{Acknowledgements}
We want to thank the Swedish Childhood Diabetes Study Group (co-ordinator Gisela Dahlquist, Ume{\aa} University) and Kerstin V\"annman for valuable discussions. 
This work was supported by the Swedish Research Council [grant numbers 0735 and 07531]; Riksbankens Jubileumsfond [grant number P11-0814:1]; and the Ume{\aa} node of the Swedish Initiative for Research on Microdata in Social and Medical Sciences funded by the Swedish Research Council. 
The simulations where run on facilities made available by the High Performance Computing Center North (HPC2N) at Ume{\aa} University.

\bibliography{Referenser}

\appendix
\section{Assumptions for existence and identification of the covariate subsets}\label{cov}

Under assumption A.2 and ${P}(X ) > 0$,  
there exists a unique minimal set $X_T \subseteq  X$ such that P$(T \mid  {X}) =  {P}(T \mid  {X}_T)$ \cite[Lemma A2]{XRW2007}.
Moreover, under A.2 and, for $t=0,1$, $ {P}(Y_t , X_T) > 0$, there exists unique and minimal subsets $Q_t$, $t=0,1$ \cite[Lemma A3]{XRW2007}. 
Corresponding positivity assumptions apply for the existence, unicity and minimality of $X_1$, $X_0$, $Z_1$ and $Z_0$ \cite[Lemma A4-5]{XRW2007}.
In addition, sufficient conditions for the subsets to be identified are assumption A.1 and $ {P}(Y_t,T,X)>0$, for $t=0,1$ \cite[Prop. 9 \& 10]{XRW2007}.

\section{Tables with results from the simulation study}\label{tables}

\begin{table}[ht!]    
\caption{The proportion (\%) of times the selected subset, $\hat{\mathsf{S}}$, satisfies three conditions of unconfoundedness. Continuous covariate distribution setup and selection is performed using MCH testing. }
\centering
\begin{tabular}{cllccccccc}
 \hline
 &  Outcome  &  Satisfied & \multicolumn{3}{c}{$\mathsf{S}$ in Algorithm A} &  \multicolumn{4}{c}{ $\mathsf{S}$ in Algorithm B} \\ 
$n$ & model & condition & $ {X}_{\text{\tiny{T}}}$ & $ {Q}_0$ & $ {Q}_1$& $ {X}_0$ & $ {X}_1$ & $ {Z}_0$ & $ {Z}_1$ \\
 \hline
500  &Linear & $Y_t \perp\!\!\!\perp T \mid \hat{\mathsf{S}}$	& 100.0 & 100.0 & 100.0 	&100.0 & 100.0 & 100.0 & 99.9 \\
     && $\mathsf{S}  \subseteq \hat{\mathsf{S}}$     		& 100.0 & 93.6 & 92.9 	& 100.0 & 100.0 & 77.2 & 78.7 \\
     && $\mathsf{S}  = \hat{\mathsf{S}}$             		& 55.6 & 18.1 & 19.8 	&  26.7 &  31.4 & 33.5 & 38.4   \\
\cline{2-10}
  &Binary & $Y_t \perp\!\!\!\perp T \mid \hat{\mathsf{S}}$	& 100.0 & 99.6 & 91.0 & 100.0 & 99.4 & 100.0 & 99.3  \\
     && $\mathsf{S}  \subseteq \hat{\mathsf{S}}$     		& 100.0 & 89.2 & 81.5 & 100.0 & 98.5  & 89.0 & 90.6 \\
     && $\mathsf{S}  = \hat{\mathsf{S}}$             		& 54.8 & 55.2 & 50.2  & 56.6 & 58.2 &  55.9 &  54.8\\
\cline{2-10}
&Nonlinear  & $Y_t \perp\!\!\!\perp T \mid \hat{\mathsf{S}}$	& 100.0 & 99.9 & 96.5	&  99.9 & 94.2 & 99.9 & 94.2  \\
     && $\mathsf{S}  \subseteq \hat{\mathsf{S}}$     		& 100.0 & 92.6 & 90.9	& 99.9 & 94.2 & 77.5 & 74.8  \\
     && $\mathsf{S}  = \hat{\mathsf{S}}$             		& 57.6 & 17.9 & 18.2 	& 27.0 & 24.6 & 35.3 & 30.0 \\
\cline{2-10}
  &Nonlinear & $Y_t \perp\!\!\!\perp T \mid \hat{\mathsf{S}}$	&  100.0 & 97.8 &  99.0  & 97.9 &  98.4  & 97.9  & 98.4 \\
  &Non-constant& $\mathsf{S}  \subseteq \hat{\mathsf{S}}$     	&  100.0 &  91.0 &  92.9  & 97.7 &  98.4  & 78.1 &  77.1 \\
     && $\mathsf{S}  = \hat{\mathsf{S}}$             		   	& 59.7  & 16.4  & 20.9 &  29.8  & 26.8  & 33.6 &  33.9 \\
\cline{2-10}
  &Nonlinear & $Y_t \perp\!\!\!\perp T \mid \hat{\mathsf{S}}$	& 100.0  & 96.3  & 97.1  & 99.3  & 95.6  & 99.2  & 95.6  \\
  &Non-constant& $\mathsf{S}  \subseteq \hat{\mathsf{S}}$     	&  100.0 &  89.2  & 90.5  & 98.9  & 95.6  & 78.1 &  76.7 \\
  & More correlation& $\mathsf{S}  = \hat{\mathsf{S}}$              & 56.0 &  16.6  & 19.3  & 30.4  & 26.2  & 29.8  & 27.4 \\
 \hline
1000  &Linear & $Y_t \perp\!\!\!\perp T \mid \hat{\mathsf{S}}$	&100.0 & 100.0 & 100.0 	& 100.0 & 100.0 & 100.0 & 100.0\\
    &   & $\mathsf{S}  \subseteq \hat{\mathsf{S}}$        	& 100.0 &  93.7 &  93.4 	& 100.0 & 100.0 & 80.0 & 78.5   \\
    & & $\mathsf{S} = \hat{\mathsf{S}}$                 		& 60.0 &  20.8 &  19.6 	&  30.3 &  29.8 & 35.1 & 34.5    \\
\cline{2-10}
  &Binary & $Y_t \perp\!\!\!\perp T \mid \hat{\mathsf{S}}$	& 100.0 &  100.0 &  99.4 &  100.0 &  100.0 &  100.0 &  100.0 \\
     && $\mathsf{S}  \subseteq \hat{\mathsf{S}}$     		& 100.0  &  90.9 &  90.7 &  100.0 &  100.0 &   90.7 &   90.6 \\
     && $\mathsf{S}  = \hat{\mathsf{S}}$             		& 57.9  &  56.9 &  57.6 &   58.9 &   57.3 &   55.2 &   56.9  \\
\cline{2-10}
&Nonlinear  & $Y_t \perp\!\!\!\perp T \mid \hat{\mathsf{S}}$	& 100.0 & 100.0 & 99.6 	& 100.0 & 99.7 & 100.0 & 99.7 \\
    && $\mathsf{S}  \subseteq \hat{\mathsf{S}}$        	& 100.0 &  92.7 & 93.2 	& 100.0 & 99.7 &  79.9 & 80.0  \\
     && $\mathsf{S} = \hat{\mathsf{S}}$                 		& 58.4 &  19.9 & 20.6 	&  29.7 & 31.0 &  36.6 & 36.8   \\
\cline{2-10}
  &Nonlinear & $Y_t \perp\!\!\!\perp T \mid \hat{\mathsf{S}}$	& 100.0 &  100.0 &  100.0 &  100.0 &  100.0 &  100.0 &  100.0 \\
  &Non-constant& $\mathsf{S}  \subseteq \hat{\mathsf{S}}$     	& 100.0 &   93.9 &   93.3 &  100.0 &  100.0  &  81.7 &   78.3 \\
     && $\mathsf{S}  = \hat{\mathsf{S}}$             		   	&  61.4  &  18.2 &   21.9  &  31.6 &   30.3   & 37.3 &   37.0 \\
\cline{2-10}
  &Nonlinear & $Y_t \perp\!\!\!\perp T \mid \hat{\mathsf{S}}$	& 100.0  & 99.8  & 100.0  & 100.0  & 100.0  & 100.0  & 100.0 \\
  &Non-constant& $\mathsf{S}  \subseteq \hat{\mathsf{S}}$     	& 100.0  & 93.8   & 94.8  & 100.0  & 100.0   & 78.5   & 78.7  \\
  &More correlation& $\mathsf{S}  = \hat{\mathsf{S}}$              & 59.2  & 19.1   & 19.4   & 30.1   & 28.8   & 29.4   & 28.7 \\
 \hline
\end{tabular}
\label{selectCont}
\end{table}

\begin{table}[ht!]    
\caption{The proportion (\%) of times the selected subset, $\hat{\mathsf{S}}$, satisfies three conditions of unconfoundedness. Discrete covariate distribution setup and selection performed using kernel smoothing.}
\centering
\begin{tabular}{cllccccccc}
 \hline
 &  Outcome  &  Satisfied & \multicolumn{3}{c}{$\mathsf{S}$ in Algorithm A} &  \multicolumn{4}{c}{ $\mathsf{S}$ in Algorithm B} \\ 
$n$ & model & condition & $ {X}_{\text{\tiny{T}}}$ & $ {Q}_0$ & $ {Q}_1$& $ {X}_0$ & $ {X}_1$ & $ {Z}_0$ & $ {Z}_1$ \\
 \hline
500 &Linear  & $Y_t \perp\!\!\!\perp T \mid \hat{\mathsf{S}}$	& 100.0 & 99.8 & 99.6 	&100.0 & 100.0 & 99.1 & 99.6 \\
     && $\mathsf{S}  \subseteq \hat{\mathsf{S}}$     		& 100.0 & 86.3 & 86.1 	& 100.0 & 100.0 & 87.9 & 86.9 \\
     && $\mathsf{S}  = \hat{\mathsf{S}}$             		& 65.0 & 49.2 & 45.9 	&  70.8 &  70.8 & 57.2 & 58.0   \\
     \cline{2-10}
  &Binary & $Y_t \perp\!\!\!\perp T \mid \hat{\mathsf{S}}$	& 100.0 &  97.4  & 86.3 &  100.0  & 98.2 &  99.2  & 97.4 \\
     && $\mathsf{S}  \subseteq \hat{\mathsf{S}}$     		& 100.0 &  85.3  & 73.9 &   99.9  & 96.4  & 73.6  & 70.5 \\
     && $\mathsf{S}  = \hat{\mathsf{S}}$             		& 62.6  & 49.8 &  36.6   & 36.6 &  25.7  & 34.6 &  28.5 \\
\cline{2-10}
&Nonlinear  & $Y_t \perp\!\!\!\perp T \mid \hat{\mathsf{S}}$	& 99.9 & 99.0 & 97.5 & 100.0 & 100.0 & 99.2 & 99.3   \\
     && $\mathsf{S}  \subseteq \hat{\mathsf{S}}$     		& 99.8 & 84.6 & 83.7 & 100.0 & 100.0 & 88.3 & 90.9  \\
     && $\mathsf{S}  = \hat{\mathsf{S}}$             		& 60.8 & 50.1 & 42.5 & 71.8 & 73.1 & 56.9 & 58.3  \\
  \cline{2-10}
  &Nonlinear & $Y_t \perp\!\!\!\perp T \mid \hat{\mathsf{S}}$	& 100.0 &  99.1 &  98.4  & 100.0 &  100.0 &  99.8 &  99.7 \\
  &Non-constant& $\mathsf{S}  \subseteq \hat{\mathsf{S}}$     	& 99.9 &  85.4  & 84.5 &  100.0  & 100.0 &  89.0  & 87.3  \\
     && $\mathsf{S}  = \hat{\mathsf{S}}$             		   	&  62.0 &  53.3 &  45.7 &   75.4   & 71.2 &  61.9  & 59.7 \\
\cline{2-10}
  &Nonlinear & $Y_t \perp\!\!\!\perp T \mid \hat{\mathsf{S}}$	& 100.0  & 99.8  & 99.7  & 100.0  & 100.0  & 99.6  & 99.4 \\
  &Non-constant& $\mathsf{S}  \subseteq \hat{\mathsf{S}}$     	& 99.7  & 81.2  & 82.3  & 100.0  & 100.0  & 89.1  & 86.9 \\
  &More correlation& $\mathsf{S}  = \hat{\mathsf{S}}$             	& 52.0  & 46.0  & 40.5   & 63.8   & 62.6  & 52.6  & 49.3 \\
 \hline
1000 &Linear  & $Y_t \perp\!\!\!\perp T \mid \hat{\mathsf{S}}$	& 100.0 & 99.7 & 99.5 &100.0  & 100.0 & 99.9 & 99.9   \\
    && $\mathsf{S}  \subseteq \hat{\mathsf{S}}$        	& 100.0 & 87.6 & 88.1 &100.0 &100.0 &93.3 & 91.8  \\
    && $\mathsf{S} = \hat{\mathsf{S}}$                 		& 73.8 & 51.4 & 51.9 & 83.9 & 82.3 & 68.7 & 68.5  \\
     \cline{2-10}
  &Binary & $Y_t \perp\!\!\!\perp T \mid \hat{\mathsf{S}}$	& 100.0 &  99.9 &  98.2 &  100.0  & 100.0 &  100.0  & 100.0 \\
     && $\mathsf{S}  \subseteq \hat{\mathsf{S}}$     		& 100.0 &  90.3 &  87.1  & 100.0  & 100.0  &  76.5  &  79.1 \\
     && $\mathsf{S}  = \hat{\mathsf{S}}$             		&  78.1  & 62.6 &  54.3   & 44.8  &  39.3  &  44.3  &  43.6 \\
\cline{2-10}
&Nonlinear  &$Y_t \perp\!\!\!\perp T \mid \hat{\mathsf{S}}$	& 100.0 &100.0 &100.0& 100.0 &100.0& 100.0& 100.0\\
  && $\mathsf{S}  \subseteq \hat{\mathsf{S}}$        	&  100.0 & 87.5 & 87.7 &100.0 &100.0&  93.0 & 92.2 \\
  && $\mathsf{S} = \hat{\mathsf{S}}$                 		&  72.8 & 55.8 & 53.6  &84.2 & 83.5 & 67.6&  69.4 \\
  \cline{2-10}
  &Nonlinear & $Y_t \perp\!\!\!\perp T \mid \hat{\mathsf{S}}$	& 100.0 &  100.0  & 100.0 &  100.0  & 100.0 &  100.0 &  100.0 \\
  &Non-constant& $\mathsf{S}  \subseteq \hat{\mathsf{S}}$     	& 100.0   & 85.9  &  87.2  & 100.0  & 100.0 &   91.6  &  92.0 \\
     && $\mathsf{S}  = \hat{\mathsf{S}}$             		   	& 74.1   & 59.2  &  54.8   & 82.2  &  82.5  &  70.3  &  71.0 \\
\cline{2-10}
  &Nonlinear & $Y_t \perp\!\!\!\perp T \mid \hat{\mathsf{S}}$	& 100.0 &  100.0 &  100.0 &  100.0 &  100.0  & 100.0  & 100.0 \\
  &Non-constant& $\mathsf{S}  \subseteq \hat{\mathsf{S}}$     	& 100.0 &   84.0 &   83.7 &  100.0  & 100.0  &  92.8   & 91.5 \\
  &More correlation& $\mathsf{S}  = \hat{\mathsf{S}}$              & 65.0   & 55.4   & 52.3  &  80.1  &  76.7  &  65.3   & 62.4 \\
 \hline
\end{tabular}
\label{selectDisc}
\end{table}

\begin{table}[ht!]    
\caption{The proportion (\%) of times the selected subset, $\hat{\mathsf{S}}$, satisfies three conditions of unconfoundedness. Mixed covariate distribution setup and selection performed using kernel smoothing. }
\centering
\begin{tabular}{cllccccccc}
 \hline
 &  Outcome  &  Satisfied & \multicolumn{3}{c}{$\mathsf{S}$ in Algorithm A} &  \multicolumn{4}{c}{ $\mathsf{S}$ in Algorithm B} \\ 
$n$ & model & condition & $ {X}_{\text{\tiny{T}}}$ & $ {Q}_0$ & $ {Q}_1$& $ {X}_0$ & $ {X}_1$ & $ {Z}_0$ & $ {Z}_1$ \\
 \hline
500 & Linear  & $Y_t \perp\!\!\!\perp T \mid \hat{\mathsf{S}}$& 100.0 & 100.0 & 99.9 	&100.0 & 100.0 & 98.1 & 98.8 \\
     && $\mathsf{S}  \subseteq \hat{\mathsf{S}}$     			& 99.9 & 85.9 & 85.4 	& 100.0 & 100.0 & 91.3 & 93.3 \\
     && $\mathsf{S}  = \hat{\mathsf{S}}$             			& 39.3 & 23.7 & 22.5 	&  62.0 &  61.6 & 30.7 & 30.9   \\
 \cline{2-10}
  &Binary & $Y_t \perp\!\!\!\perp T \mid \hat{\mathsf{S}}$	& 99.9 &  93.2 &  88.5 &  99.9 &  99.4 &  98.5 &  97.6 \\
     && $\mathsf{S}  \subseteq \hat{\mathsf{S}}$     		& 99.7 &  79.5 &  75.9 &  99.3 &  98.8 &  73.2 &  74.1 \\
     && $\mathsf{S}  = \hat{\mathsf{S}}$             		& 34.8  & 16.8 &  14.4 &  22.8 &  21.4 &  14.5 &  14.2 \\
\cline{2-10}
&Nonlinear  & $Y_t \perp\!\!\!\perp T \mid \hat{\mathsf{S}}$	& 99.8 & 84.2 & 85.1 	& 100.0& 99.9 & 97.4 & 97.8  \\
     && $\mathsf{S}  \subseteq \hat{\mathsf{S}}$     			& 99.8 & 70.5 & 74.0 	& 100.0 & 99.9 & 90.0 & 90.0  \\
     && $\mathsf{S}  = \hat{\mathsf{S}}$             			& 33.2 & 12.6 & 11.7	& 57.3 & 55.4 & 28.1 & 26.3  \\
  \cline{2-10}
  &Nonlinear & $Y_t \perp\!\!\!\perp T \mid \hat{\mathsf{S}}$	&  99.9 &  83.6 &  86.5 &  99.9 &  99.9 &  98.8 &  98.7 \\
  &Non-constant& $\mathsf{S}  \subseteq \hat{\mathsf{S}}$     	& 99.6 &  71.4 &  73.4  & 99.9 &  99.9 &  90.6  & 90.6   \\
     && $\mathsf{S}  = \hat{\mathsf{S}}$             		   	& 34.7  & 12.2 &  13.5  & 55.8 &   55.7 &  27.3  & 26.9 \\
\cline{2-10}
  &Nonlinear & $Y_t \perp\!\!\!\perp T \mid \hat{\mathsf{S}}$	&  99.8 &  83.0  & 93.7 &  100.0  & 100.0 &  97.3  & 97.4 \\
  &Non-constant& $\mathsf{S}  \subseteq \hat{\mathsf{S}}$     	&  99.4  & 68.7  & 78.7  & 100.0  & 100.0  & 89.8  & 88.8  \\
  &More correlation& $\mathsf{S}  = \hat{\mathsf{S}}$              &  33.2  & 11.8 &  13.2   & 56.2 &   54.5  & 27.4 &  28.9 \\
 \hline
1000 & Linear  & $Y_t \perp\!\!\!\perp T \mid \hat{\mathsf{S}}$	& 100.0 & 100.0 &99.9 &100.0 & 100.0& 99.9 &100.0 \\
   && $\mathsf{S}  \subseteq \hat{\mathsf{S}}$        	&  100.0 & 88.5 &88.1 &100.0 &100.0& 96.4  &97.6  \\
   && $\mathsf{S} = \hat{\mathsf{S}}$                 		&   49.5  &28.5& 28.5 & 66.0 & 68.5 &37.0  &40.1 \\
     \cline{2-10}
  &Binary & $Y_t \perp\!\!\!\perp T \mid \hat{\mathsf{S}}$	&  100.0 &  99.8  & 98.4 &  100.0 &  100.0 &  100.0  & 100.0 \\
     && $\mathsf{S}  \subseteq \hat{\mathsf{S}}$     		&  100.0  & 90.6  & 88.9  & 100.0 &   99.9  &  79.4 &   83.0  \\
     && $\mathsf{S}  = \hat{\mathsf{S}}$             		&  45.9  & 23.4 &  20.9   & 31.9 &   32.6   & 18.9 &   20.6 \\
\cline{2-10}
&Nonlinear  &$Y_t \perp\!\!\!\perp T \mid \hat{\mathsf{S}}$	& 100.0 & 96.1 &96.6 &100.0 &100.0& 100.0& 100.0 \\
   && $\mathsf{S}  \subseteq \hat{\mathsf{S}}$        	& 100.0 &86.4  &87.2 &100.0& 100.0  & 95.9 &  95.8  \\
   && $\mathsf{S} = \hat{\mathsf{S}}$                 		&  45.8 &20.6 & 20.4  &61.7&  60.8 & 33.2 & 31.3  \\
  \cline{2-10}
  &Nonlinear & $Y_t \perp\!\!\!\perp T \mid \hat{\mathsf{S}}$	& 100.0  & 96.3  & 97.0 &  100.0  & 100.0  & 100.0  & 99.9 \\
  &Non-constant& $\mathsf{S}  \subseteq \hat{\mathsf{S}}$     	& 100.0  & 88.1  & 87.0  & 100.0  & 100.0   & 96.1 &  96.8 \\
     && $\mathsf{S}  = \hat{\mathsf{S}}$             		   	& 44.5  & 19.8  & 21.0   & 63.5  &  62.1   & 32.4 &  31.6  \\
\cline{2-10}
  &Nonlinear & $Y_t \perp\!\!\!\perp T \mid \hat{\mathsf{S}}$	&  100.0 &  92.4  & 99.3  & 100.0  & 100.0  & 100.0  & 99.9 \\
  &Non-constant& $\mathsf{S}  \subseteq \hat{\mathsf{S}}$     	&  100.0 &  80.9  & 88.3  & 100.0  & 100.0  & 96.1 &  95.6 \\
  &More correlation& $\mathsf{S}  = \hat{\mathsf{S}}$              &  47.9 &  20.9 &  20.4   & 63.1  &  62.1  & 34.4 &  32.3 \\
 \hline
\end{tabular}
\label{selectMix}
\end{table}

\begin{table}[ht!]   
\caption{Results from the \emph{continuous covariate distribution} setup, with $n=500$, when estimating $\beta$ using matching on the vector norm and the propensity score and IPW. $m$ is the median cardinality of the conditioning set.}
\centering
\begin{tabular}{llcccccccccc}
 \hline
 Outcome&  & \multicolumn{3}{c}{Vector norm} & \multicolumn{3}{c}{Propensity Score}  & \multicolumn{3}{c}{IPW} &  \\ 
 model   &  Set       & Bias  & SD & MSE   & Bias  & SD & MSE & Bias  & SD & MSE & $m$  \\
 \hline
Linear & $ {X}$ 		& 1.342 & 0.251 & 1.865		& 0.175 & 0.626 & 0.423  		& 0.010 & 1.522 & 2.315 		& 10\\
 & ${\hat{ {X}}_{\text{\tiny{T}}}}$	& 1.066 & 0.385 & 1.285		& 0.175 & 0.715 & 0.542  		& 0.044 & 1.362 & 1.856 		& 5 \\
  & ${\hat{ {Q}}}$ 				& 0.839 & 0.411 & 0.873		& 0.132 & 0.588 & 0.363  		& 0.051 & 1.123 &  1.264		& 5 \\
& ${\hat{ {X}}_{\text{\tiny{Y}}}}$ 	& 1.027 & 0.274 & 1.130		& 0.087 & 0.427 & 0.190  		& 0.039 & 0.671 & 0.452		& 7  \\
 & ${\hat{ {Z}}}$ 				& 0.670 & 0.376 & 0.631		& 0.109 & 0.468 & 0.231  		& 0.055 & 0.678 & 0.463 		& 4  \\
 \hline
 Binary & $ {X}$ 				& 0.099 & 0.040 & 0.011 			& 0.010 & 0.077 & 0.006 			& -0.002 & 0.168 & 0.028 		& 10  \\
 & ${\hat{ {X}}_{\text{\tiny{T}}}}$	& 0.076 & 0.048 & 0.008 			& 0.010 & 0.077 & 0.006 			& -0.002 & 0.170 & 0.029 		& 5 \\
  &  ${\hat{ {Q}}}$  			& 0.040 & 0.047 & 0.004 			& 0.006 & 0.056 & 0.003 			& 0.004 & 0.062 & 0.004 			& 4  \\
 & ${\hat{ {X}}_{\text{\tiny{Y}}}}$ 	& 0.064 & 0.038 & 0.006 			& 0.004 & 0.049 & 0.002 			& 0.004 & 0.053 & 0.003 			& 6  \\
  &${\hat{ {Z}}}$	 			& 0.040 & 0.046 & 0.004 			& 0.003 & 0.051 & 0.003 			& 0.004 & 0.055 & 0.003 			& 4  \\
 \hline
 Nonlinear & $ {X}$ 			& 0.762 & 0.305 & 0.673		& 0.122 & 0.947 & 0.911		 & 0.025 & 1.317 & 1.735 		&10 \\
 & ${\hat{ {X}}_{\text{\tiny{T}}}}$	& 0.657 & 0.484 & 0.666		& 0.146 & 1.030 & 1.082 		 & 0.040 & 1.333 & 1.778 		&5\\
 &  ${\hat{ {Q}}}$  			& 0.524 & 0.495 & 0.519		& 0.117 & 0.882 & 0.791		 & 0.065 & 0.960 & 0.926 		&5\\
 & ${\hat{ {X}}_{\text{\tiny{Y}}}}$ 	& 0.623 & 0.305 & 0.481		& 0.098 & 0.708 & 0.510		 & 0.037 & 0.710 & 0.505 		&7\\
 &${\hat{ {Z}}}$	 			& 0.421 & 0.467 & 0.395		& 0.087 & 0.757 & 0.580 		 & 0.040 & 0.792 & 0.629		 &5\\
 \hline
Nonlinear & $ {X}$ 		 	& 1.062 & 0.361 & 1.258 			& 0.091 & 1.198 & 1.443 			& -0.089 & 2.049 & 4.208 	& 10  \\
 Nonconstant & ${\hat{ {X}}_{\text{\tiny{T}}}}$	& 0.859 & 0.552 & 1.042 			& 0.143 & 1.233 & 1.540 			& -0.032 & 1.920 & 3.687 	& 5 \\
  & ${\hat{ {Q}}}$ 						& 0.690 & 0.591 & 0.826 			& 0.124 & 1.092 & 1.209 			& 0.067 & 1.365 & 1.868 		& 5 \\
 & ${\hat{ {X}}_{\text{\tiny{Y}}}}$ 			& 0.821 & 0.360 & 0.804 			& 0.108 & 0.967 & 0.946 			& 0.025 & 1.346 & 1.812 		& 7 \\
 & ${\hat{ {Z}}}$ 						& 0.559 & 0.552 & 0.618 			& 0.111 & 0.977 & 0.967 			& 0.041 & 1.373 & 1.887 		&  5 \\
 \hline
 Nonlinear & $ {X}$ 					& 1.182 & 0.343 & 1.515 			& 0.270 & 1.301 & 1.765 			& 0.139 & 2.188 & 4.808 	& 10 \\
 Nonconstant& ${\hat{ {X}}_{\text{\tiny{T}}}}$	& 0.920 & 0.599 & 1.205 			& 0.198 & 1.374 & 1.926 			& 0.145 & 2.171 & 4.734 	& 5 \\
  &  ${\hat{ {Q}}}$  			& 0.728 & 0.627 & 0.923 			& 0.177 & 1.238 & 1.564 			& 0.076 & 1.813 & 3.291 	& 5\\
 More & ${\hat{ {X}}_{\text{\tiny{Y}}}}$ 			& 0.908 & 0.358 & 0.952 			& 0.189 & 1.062 & 1.163 			& 0.146 & 1.395 & 1.968 	& 7 \\
 correlation&${\hat{ {Z}}}$	 					& 0.647 & 0.582 & 0.757 			& 0.174 & 1.094 & 1.226 			& 0.125 & 1.384 & 1.930 	& 5 \\
 \hline
\end{tabular}
\label{estimateCont}
\end{table}

\begin{table}[ht!]   
\caption{Results from the \emph{discrete covariate distribution} setup, with $n=500$,  when estimating $\beta$ using matching on the vector norm and the propensity score and IPW. $m$ is the median cardinality of the conditioning set.}
\centering
\begin{tabular}{llcccccccccc}		
 \hline
 Outcome&  & \multicolumn{3}{c}{Vector norm} & \multicolumn{3}{c}{Propensity Score}  & \multicolumn{3}{c}{IPW} &  \\ 
 model   &  Set       & Bias  & SD & MSE   & Bias  & SD & MSE & Bias  & SD & MSE & $m$  \\
 \hline
 Linear& $ {X}$ 			& 1.161 & 0.238 & 1.406		& 0.109 & 0.637&  0.417 		& 0.037  & 3.129  & 9.793 	& 10\\
&  ${\hat{ {X}}_{\text{\tiny{T}}}}$	& 0.482 & 0.493 & 0.475		& 0.099 & 0.712&  0.517		& 0.005  & 2.859  & 8.173 	& 5 \\
 & ${\hat{ {Q}}}$ 				& 0.120 & 0.461 & 0.227		& 0.077 & 0.498 & 0.254 		& 0.115  & 1.100  & 1.223		& 4 \\
& ${\hat{ {X}}_{\text{\tiny{Y}}}}$ 	& 0.231 & 0.250 & 0.116		& 0.015 & 0.281 & 0.079 		& 0.038  & 0.708  & 0.503		& 5  \\
 & ${\hat{ {Z}}}$ 				& 0.045 & 0.341 & 0.118		& 0.001 & 0.363 &  0.132		& 0.030  & 0.728  & 0.531		& 4  \\
 \hline
 Binary & $ {X}$ 				& -0.069 & 0.033 & 0.006		& -0.007 & 0.072 & 0.005		& -0.002 & 0.102 & 0.010 		& 10 \\
 & ${\hat{ {X}}_{\text{\tiny{T}}}}$	& -0.028 & 0.049 & 0.003		& -0.005 & 0.072 & 0.005		& -0.002 & 0.098 & 0.010 		& 5 \\
 &  ${\hat{ {Q}}}$  			& -0.009 & 0.049 & 0.002		& -0.005 & 0.055 & 0.003		& -0.006 & 0.058 & 0.003 		&  4 \\
 & ${\hat{ {X}}_{\text{\tiny{Y}}}}$ 	& -0.034 & 0.037 & 0.003		& -0.006 & 0.053 & 0.003		& -0.002 & 0.065 & 0.004 		& 7 \\
 &${\hat{ {Z}}}$	 			& -0.012 & 0.045 & 0.002		& -0.005 & 0.054 & 0.003		& -0.002 & 0.066 & 0.004 		&  5 \\
 \hline
 Nonlinear & $ {X}$ 			& 1.005 & 0.276& 1.086 		& 0.089 & 0.825 & 0.688		& 0.022 & 1.351 & 1.825  		&  10\\
 & ${\hat{ {X}}_{\text{\tiny{T}}}}$	& 0.461 & 0.600& 0.573		& 0.083 & 0.897 & 0.812	 	& 0.019 & 1.356 & 1.840 		& 5\\
 &  ${\hat{ {Q}}}$  			& 0.162 & 0.616& 0.405		& 0.094 & 0.646 &0.427	 	& 0.065 & 0.829 & 0.692 		& 4\\
 & ${\hat{ {X}}_{\text{\tiny{Y}}}}$ 	& 0.191 & 0.222& 0.086		& 0.013 & 0.328 & 0.107		& 0.024 & 0.398 & 0.159 		& 	5\\
 &${\hat{ {Z}}}$	 			& 0.032 & 0.446& 0.200		& -0.010 & 0.470 & 0.221 	& 0.012 & 0.501 & 0.252 		& 4\\
 \hline
Nonlinear & $ {X}$ 		 			& 1.214 & 0.304 & 1.566 			& 0.127 & 0.820 & 0.688 		& -0.062 & 1.752 & 3.073 		& 10  \\
Nonconstant & ${\hat{ {X}}_{\text{\tiny{T}}}}$	& 0.512 & 0.637 & 0.668 			& 0.146 & 0.922 & 0.872 		& -0.065 & 1.794 & 3.224 		&  5\\
  & ${\hat{ {Q}}}$ 						& 0.136 & 0.617 & 0.399 			& 0.087 & 0.651 & 0.432 		& 0.077  & 0.886 & 0.790 		& 4 \\
 & ${\hat{ {X}}_{\text{\tiny{Y}}}}$ 			& 0.225 & 0.268 & 0.122 			& 0.013 & 0.345 & 0.119 		& 0.027  & 0.528 & 0.280 		& 5 \\
 & ${\hat{ {Z}}}$ 						& 0.048 & 0.466 & 0.219 			& 0.015 & 0.495 & 0.245 		& 0.031  & 0.607 & 0.370 		&  4 \\
 \hline
 Nonlinear & $ {X}$ 					& 1.188 & 0.316 & 1.511 			& 0.093 & 0.784 & 0.624 		& -0.110 & 2.171 & 4.727 		& 10	\\
 Nonconstant& ${\hat{ {X}}_{\text{\tiny{T}}}}$	& 0.488 & 0.561 & 0.552 			& 0.038 & 0.813 & 0.663 		& -0.128 & 2.119 & 4.508 		&  5\\
 &  ${\hat{ {Q}}}$  			& 0.104 & 0.516 & 0.277 			& 0.038 & 0.577 & 0.334 		& 0.056  & 0.934 & 0.876 		&   4 \\
 More & ${\hat{ {X}}_{\text{\tiny{Y}}}}$ 			& 0.231 & 0.298 & 0.142 			& 0.014 & 0.362 & 0.131 		& 0.046  & 0.609 & 0.373 		& 6 \\
 correlation&${\hat{ {Z}}}$	 					& 0.062 & 0.410 & 0.172 			& 0.003 & 0.455 & 0.207 		& 0.038  & 0.654 & 0.430 		& 4 \\
 \hline
\end{tabular}
\label{estimateDisc}
\end{table}

\begin{table}[ht!]   
\caption{Results from the \emph{mixed covariate distribution} setup, with $n=500$,  when estimating $\beta$ using matching on the vector norm and the propensity score and IPW. $m$ is the median cardinality of the conditioning set.}
\centering
\begin{tabular}{llcccccccccc}
 \hline
 Outcome&  & \multicolumn{3}{c}{Vector norm} & \multicolumn{3}{c}{Propensity Score}  & \multicolumn{3}{c}{IPW} &  \\ 
 model   &  Set       & Bias  & SD & MSE   & Bias  & SD & MSE & Bias  & SD & MSE & $m$  \\
 \hline
Linear & $ {X}$ 			& 1.292 & 0.249 & 1.732		& 0.157 & 0.617&  0.405 		& 0.016 & 1.776 & 3.154 		& 10\\
 & ${\hat{ {X}}_{\text{\tiny{T}}}}$	& 0.886 & 0.428 & 0.969		& 0.174 & 0.721&  0.550		& $<$0.001 & 1.816 & 3.296 	&  6 \\
 & ${\hat{ {Q}}}$ 				& 0.502 & 0.453 & 0.457		& 0.131 & 0.519 & 0.286 		& 0.126 & 0.899 & 0.824 		&  4 \\
 & ${\hat{ {X}}_{\text{\tiny{Y}}}}$ 	& 0.639 & 0.302 & 0.499		& 0.073 & 0.341 & 0.122 		& 0.075 & 0.604 & 0.371 		&  6  \\
& ${\hat{ {Z}}}$ 				& 0.391 & 0.393 & 0.307		& 0.086 & 0.406 &  0.173		& 0.077 & 0.639 & 0.415 		&  4  \\
 \hline
 Binary & $ {X}$ 				& -0.085 & 0.037 & 0.009		& -0.011 & 0.074 & 0.006 	& -0.007 & 0.100 & 0.010 		& 10  \\
 & ${\hat{ {X}}_{\text{\tiny{T}}}}$	& -0.057 & 0.048 & 0.006		& -0.009 & 0.074 & 0.006 	& -0.008 & 0.095 & 0.009 		& 6 \\
  &  ${\hat{ {Q}}}$  			& -0.036 & 0.049 & 0.004		& -0.008 & 0.061 & 0.004 	& -0.010 & 0.081 & 0.007 		& 5  \\
 & ${\hat{ {X}}_{\text{\tiny{Y}}}}$ 	& -0.060 & 0.041 & 0.005		& -0.006 & 0.054 & 0.003 	& -0.011 & 0.061 & 0.004 		& 7  \\
 &${\hat{ {Z}}}$	 			& -0.037 & 0.046 & 0.003		& -0.009 & 0.055 & 0.003 	& -0.011 & 0.061 & 0.004 		& 5  \\
 \hline
Nonlinear & $ {X}$ 			& 0.673 & 0.282 & 0.533 		& 0.033 & 0.740 & 0.549		& -0.021 & 1.395 & 1.945 		&  10\\
 & ${\hat{ {X}}_{\text{\tiny{T}}}}$	& 0.495 & 0.495 & 0.491 		& 0.029 & 0.835 & 0.699		& -0.027 & 1.555 & 2.419		 &  6\\
 &  ${\hat{ {Q}}}$  			& 0.291 & 0.500 & 0.334 		& 0.043 & 0.670 & 0.490		& -0.005 & 0.890 & 0.792		 &  5\\
& ${\hat{ {X}}_{\text{\tiny{Y}}}}$ 	& 0.335 & 0.260 & 0.180 		& 0.016 & 0.476 & 0.227		& 0.030 & 0.469 & 0.221		 &  6\\
 &${\hat{ {Z}}}$	 			& 0.196 & 0.435 & 0.228 		& 0.032 & 0.555 & 0.309		& 0.031 & 0.560 & 0.314		 &  5\\
 \hline
Nonlinear & $ {X}$ 		 			& 0.689 & 0.281 & 0.554 			& 0.087 & 0.773 & 0.606 			&  $<$0.001 & 1.295 & 1.676 		& 10  \\
 Nonconstant & ${\hat{ {X}}_{\text{\tiny{T}}}}$	& 0.509 & 0.518 & 0.527 			& 0.104 & 0.839 & 0.714 			& -0.003 & 1.310 & 1.717 		& 6 \\
  & ${\hat{ {Q}}}$ 						& 0.292 & 0.541 & 0.378 			& 0.066 & 0.707 & 0.504 			& 0.017 & 0.953 & 0.908 		& 5 \\
 & ${\hat{ {X}}_{\text{\tiny{Y}}}}$ 			& 0.346 & 0.258 & 0.186 			& 0.028 & 0.482 & 0.233 			& 0.013 & 0.463 & 0.214 		& 6 \\
 & ${\hat{ {Z}}}$ 						& 0.198 & 0.450 & 0.242 			& 0.025 & 0.552 & 0.305 			& 0.007 & 0.553 & 0.306 		&  4 \\
 \hline 
 Nonlinear & $ {X}$ 					&  0.679 & 0.303 & 0.553 			& 0.070 & 0.799 & 0.644 			&  $<$0.001 & 1.518 & 2.303 		& 10\\
 Nonconstant& ${\hat{ {X}}_{\text{\tiny{T}}}}$	& 0.480 & 0.516 & 0.500			& 0.047 & 0.889 & 0.792 			& 0.002 & 1.422 & 2.022 		& 6 \\
  &  ${\hat{ {Q}}}$  					& 0.259 & 0.528 & 0.346 			& 0.029 & 0.736 & 0.543 			& 0.029 & 0.847 & 0.719 		& 5 \\
 More & ${\hat{ {X}}_{\text{\tiny{Y}}}}$ 		& 0.322 & 0.271 & 0.177 			& 0.036 & 0.531 & 0.283 			& -0.003 & 0.583 & 0.340 		& 6 \\
 correlation &${\hat{ {Z}}}$	 			& 0.186 & 0.448 & 0.235 			& 0.046 & 0.590 & 0.351 			& 0.012 & 0.617 & 0.381		& 5 \\
 \hline
\end{tabular}
\label{estimateMix}
\end{table}

\begin{table}
\caption{Descriptive statistics displaying group proportions (\%) for the discrete variables and mean values from the continuous variables. Education is categorized into three levels: Compulsory Schooling (CS), Upper Secondary Education (SE) and Higher Education (HE). } 
\centering
\begin{tabular}{ l  c  c }
	 \hline
	Variable  										& Low grades 	& Higher grades  \\ 
	 \hline
	Hospitalization with acute complications				&	22.3\% 			& 	14.1\%		\\
	Female gender (\emph{sex}) 						&  	35.6\%			& 	51.2\%  \\
	Mean age at onset of T1DM	(\emph{ageO})			&	8.3 years   		& 	8.5 years  \\
	Mother's mean age at delivery (\emph{ageM})			&	27.3 years		&	28.0 years \\
	Father's mean age at delivery (\emph{ageF})			&	30.1 years		&	30.7 years \\
	Social benefits in the family (\emph{soc})	&				11.2\% 			& 	3.6\%	\\
	Single mother status		(\emph{single})		&			24.3\%   			& 	15.9\%  \\
	Education level		(CS/SE/HE)		&&	\\
	\, \, - mother 	(\emph{educM})					&  36.0\% /53.6\% /10.4\%    &	19.1\% /48.8\% /32.2\%   \\
	\, \, - father	(\emph{educF})					&  42.9\% /46.4\% /10.8\%  &	 26.7\% /45.4\% /27.9\%  \\ 
	Mean salary income  	&	\\
	\, \, - mother	(\emph{incM})					&	96 700 SEK			&	124 800 SEK	\\
	\, \, - father	(\emph{incF})					&	149 300 SEK			&	204 600 SEK  \\
	Sick pay			&\\
	\, \, - mother	(\emph{sickM})					& 32.1\%				&	33.4\% \\
	\, \, - father	(\emph{sickF})					& 32.3\%				&	25.1\% \\
	Unemployment benefits &	\\
	\, \, - mother	(\emph{unempM})					& 12.7\%				&	10.9\% \\
	\, \, - father	(\emph{unempF})					& 12.5\%				&	6.8\% \\
	Benefits for labour market policy measures &\\
	\, \, - mother	(\emph{ampolM})					& 6.1\%				&	 4.2\%     \\
	\, \, - father	(\emph{ampolF})					& 8.4\%				&	3.7\%	\\
	Early retirement and sick benefits	&\\
	\, \, - mother	(\emph{erM})						& 7.6\%				&	2.7\% \\
	\, \, - father	(\emph{erF})						& 6.3\%				&	2.6\% \\
	Retirement pension				&	\\
	\, \, - mother		(\emph{retM})					& 0.8\%				&	0.8\% \\
	\, \, - father		(\emph{retF})					& 1.2\%				&	1.7\% \\
	Occupational injury annuity 	&	\\ 
	\, \, - mother			(\emph{injM})				& 1.6\%				&    0.7\% \\
	\, \, - father			(\emph{injF})				& 4.7\%				&    1.5\% \\	
	 \hline
\end{tabular}
\label{desc}
\end{table}

\end{document}